\documentclass[a4paper,11pt]{article}
\pdfoutput=1
\usepackage{jcappub} 

\usepackage{subcaption}
\usepackage{multirow}
\usepackage{amsmath}
\usepackage{svn-multi}
\usepackage{xspace}
\usepackage{flafter}
\usepackage{color}
\usepackage{soul}
\usepackage{graphicx}
\usepackage{graphics,epsfig}

\newcommand {\gtsim}  {\lower.5ex\hbox{$\; \buildrel > \over \sim \;$}} 
\newcommand {\ltsim}  {\lower.5ex\hbox{$\; \buildrel < \over \sim \;$}}
\newcommand{\twof}{$2\kern -0.1em f$~}
\newcommand{\fourf}{$4\kern -0.1em f$~}
\newcommand{\oneoverf}{$1/\kern -0.1em f$~}

\svnid{$Id:$}

\definecolor{orange}{rgb}{1,0.5,0.2}

\newcommand  \beq    {\begin{equation}}

\newcommand  \eeq    {\end{equation}}

\newcommand{\lap} {$\stackrel{<}{_\sim}$}

\title{Results from the Atacama B-mode Search (ABS) Experiment}
\dedicated{Draft compiled on:  \today}

\author[a,b,c,1]{Akito Kusaka,\note{Corresponding author.}}
\author[a,d]{John Appel,}
\author[a,e]{Thomas Essinger-Hileman,}
\author[f]{James A. Beall,}
\author[g]{Luis E. Campusano,}
\author[h]{Hsiao-Mei Cho,}
\author[a]{Steve K. Choi,}
\author[a]{Kevin Crowley,}
\author[f]{Joseph W. Fowler,}
\author[i,j]{Patricio Gallardo,}
\author[k]{Matthew Hasselfield,}
\author[f]{Gene Hilton,}
\author[a]{Shuay-Pwu P. Ho,}
\author[h,l]{Kent Irwin,}
\author[a]{Norman Jarosik,}
\author[i]{Michael D. Niemack,}
\author[a,m]{Glen W. Nixon,}
\author[n]{Michael  Nolta,}
\author[a]{Lyman A. Page Jr,}
\author[o]{Gonzalo A. Palma,}
\author[a]{Lucas Parker,}
\author[g,p]{Srinivasan Raghunathan,}
\author[f]{Carl D. Reintsema,}
\author[q,r]{Jonathan Sievers,}
\author[a,s]{Sara M. Simon,}
\author[a]{Suzanne T. Staggs,}
\author[a]{Katerina Visnjic,}
\author[h]{Ki-Won Yoon}

\affiliation[a]{\small Dept. of Physics, Princeton University, Jadwin Hall, Princeton, NJ 08544, USA}
\affiliation[b]{\small Physics Division, Lawrence Berkeley National Laboratory, Berkeley, CA 94720, USA  }
\affiliation[c]{\small Dept. of Physics, University of Tokyo, Tokyo 113-0033, Japan }
\affiliation[d]{\small Dept. of Physics \& Astronomy, Johns Hopkins University, 3400 N. Charles Street, Baltimore, MD 21218, USA}
\affiliation[e]{\small NASA, Code 665, Goddard Space Flight Center, Greenbelt, MD 20771, USA}
\affiliation[f]{\small National Institute of Standards and Technology (NIST), Boulder, Colorado 80305, USA}
\affiliation[g]{\small Departamento de Astronom\'{i}a, Universidad de Chile, 
Camino del Observatorio 1515, Santiago, Chile}
\affiliation[h]{\small Stanford Linear Accelerator Center,  2575 Sand Hill Rd.  Menlo Park, CA 94025, USA}
\affiliation[i]{\small Department of Physics, Cornell University, Ithaca, NY 14853, USA }
\affiliation[j]{\small Dept. of Ast. \& Astrop. Pontificia Universidad Cat\'olica de Chile,
Vicu\~na Mackenna, 4860 Casilla 306, Santiago, 22 Chile}
\affiliation[k]{\small Dept. of Ast. \& Astrop., Penn State University, 525 Davey Lab
University Park, PA 16802, USA}
\affiliation[l]{\small Department of Physics, Stanford University, Stanford, CA 94305, USA }
\affiliation[m]{\small Citadel Securities, 131 South Dearborn Street, Chicago, Illinois, USA}
\affiliation[n]{\small Canadian Institute for Theoretical Astrophysics, University of Toronto, 60 St. George Street, Toronto, ON M5S 3H8, Canada}
\affiliation[o]{\small Departamento de F\'{i}sica, FCFM, Universidad de Chile, Blanco Encalada 
2008, Santiago, Chile}
\affiliation[p]{\small School of Physics, University of Melbourne, Parkville VIC 3010, Australia}
\affiliation[q]{\small School of Chemistry and Physics, University of KwaZulu-Natal Durban, 4041, South Africa}
\affiliation[r]{\small National Institute for Theoretical Physics (NITheP), KZN node, Durban, South Africa}
\affiliation[s]{\small Department of Physics, University of Michigan, 450 Church St., Ann Arbor, Michigan, USA}

\emailAdd{akusaka@lbl.gov}
\emailAdd{jappel3@jhu.edu}
\emailAdd{thomas.m.essinger-hileman@nasa.gov}
\emailAdd{staggs@princeton.edu}

\abstract{The Atacama B-mode Search (ABS) is an experiment designed to measure cosmic microwave background (CMB) polarization at large angular scales ($\ell>40$). It operated from the ACT site at 5190~m elevation in northern Chile at 145~GHz with a net sensitivity (NEQ) of 41~$\mu$K$\sqrt{\rm s}$. It employed an 
ambient-temperature sapphire half-wave plate rotating at 2.55~Hz to modulate the incident polarization signal and reduce systematic effects. We report here on the analysis of data from a 2400 deg$^2$ patch of sky centered at declination $-42^\circ$ and right ascension $25^\circ$. We perform a blind analysis. After unblinding, we find agreement with the Planck TE and EE measurements on the same region of sky. We marginally detect polarized dust emission and give an upper limit on the tensor-to-scalar ratio of $r<2.3$ (95\% cl) with  the equivalent of 100 on-sky days of observation. We also present a new measurement of the polarization of Tau A and introduce new methods associated with HWP-based observations.}


\begin{document}
\maketitle
\flushbottom

\section{Introduction}
\label{sec:intro}

Measurements of the CMB polarization at angular scales between $1^{\circ}$ and $10^{\circ}$ ($20<\ell<200$)   have the potential to reveal primordial gravitational waves. 
A detection of them would be profound, potentially providing a glimpse of gravity operating on a quantum scale, and constraining models of the early universe. 

The influence of gravitational waves on the large angular scale CMB temperature anisotropy has been appreciated since the seminal \citet{sachs/wolfe:1967} paper. Although soon thereafter \citet{rees:1968} noted that anisotropies in the primordial CMB lead to its linear polarization, it was not until \citet{polnarev:1985} that the connection was made between gravitational waves and the generation of large angular scale polarization, the focus of our investigation. \citet{crittenden/davis/steinhardt:1993} noted that gravitational waves lead to a possibly distinctive polarized signal at large angular scales and gave the spectrum based on a Boltzmann transport code while \citet{harari/zaldarriaga:1993} gave an analytic treatment based on \citet{basko/polnarev:1980}. 

The modern framework for quantifying the CMB polarization is given in  \citet{kamionkowski/etal/1997b} and 
\citet{zaldarriaga/seljak:1997}. They present coordinate-independent frameworks for computing the linear polarization over the full sky for open, closed, and flat geometries, quantified as ``E-modes" and ``B-modes," distinguished by their global symmetry properties.  Both groups 
went on to show that gravitational waves uniquely produce B-mode polarization in the CMB at large angular scales in addition to producing E-mode polarization and temperature anisotropies \cite{kamionkowski/etal/1997a, seljak/zaldarriaga:1997}.  
 \citet{zaldarriaga/seljak:1998} showed that the same physical process that produces gravitational lensing of the temperature anisotropy results in some B-mode polarization even when the primordial polarization is purely E-mode.  Fortunately, these ``lensing" B-modes can be measured and 
largely subtracted from maps, and/or distinguished from the ``primordial" B-modes by their spectral shape.   
At the ABS sensitivity level,  the lensing B-modes are not detectable.
On the other hand,  polarized galactic foreground emission from synchrotron and dust are potential contaminants.

The current limit on primordial gravitational waves comes from the BICEP2/Keck  experiment (hereafter B2K). The B2K team reports a tensor-to-scalar ratio of $r<0.09$  at the 95\% confidence level (cl) with a pivot scale of $k=0.05$~Mpc$^{-1}$ \cite{B2K-2016}.  The limit tightens to $r<0.07$ when B2K data are combined with both CMB temperature and polarization data from Planck
(e.g., \citep{bk-planck:2015}).

The lensing B-mode has been detected over a range of angular scales by SPT \citep{hanson/etal/2013, keisler/etal:2015}, {\sc polarbear} \citep{polarbear:2014, pbear-herschel/2013, pbear-eebb:2014}, B2K \citep{bicep2_results:2014, bk-planck:2015}, Planck \citep{planck_lens:2015}, and ACT \citep{vanengelen/etal:2014, louis_ACTpol_2016}. Both synchrotron and/or dust foreground B-modes have been measured at large angular scales by Archeops \citep{benoit/etal:2004,ponthieu/etal:2005}, WMAP \citep{gold/etal:2011}, Planck \citep{planck_fg:2016}, and 
BICEP2/Keck \citep{B2K-2016}. 

ABS  is one of several instruments developed with the express purpose of detecting primordial B-modes if they are sufficiently large. Other new-millennium experiments that have published results on polarization at $\ell<100$ include POLAR \citep{keating/etal:2001},  PIQUE \citep{hedman/etal:2002}, WMAP \citep{page/etal:2007}, 
BICEP \citep{chiang/etal:2010}, QUIET \citep{quiet:2011}, B2K, Planck \citep{planck_overview:2016}, and SPT \citep{henning/etal:2017}. 

One of the unique features of ABS is its use of a rapidly rotating ambient-temperature half-wave plate (HWP) to modulate the incident polarized signal at frequencies above where atmospheric fluctuations dominate over detector noise \cite{kusaka:2014_HWP}, and above where thermal drifts  contaminate the signal.  The HWP also suppresses systematic  effects \cite{essinger-hileman/etal:2015}.
We present here limits on $r$ from two years of ABS data, with emphasis on the analysis techniques for, and benefits from, the HWP.

\section{The ABS Instrument}
\label{sec:instrument}
 
\begin{figure}[ht!]
	\centering
	\begin{subfigure}[b]{0.4\textwidth}
		\centering
		\includegraphics[height=1\textwidth, clip=true, trim=0.5in 0.75in 0in 2in ]{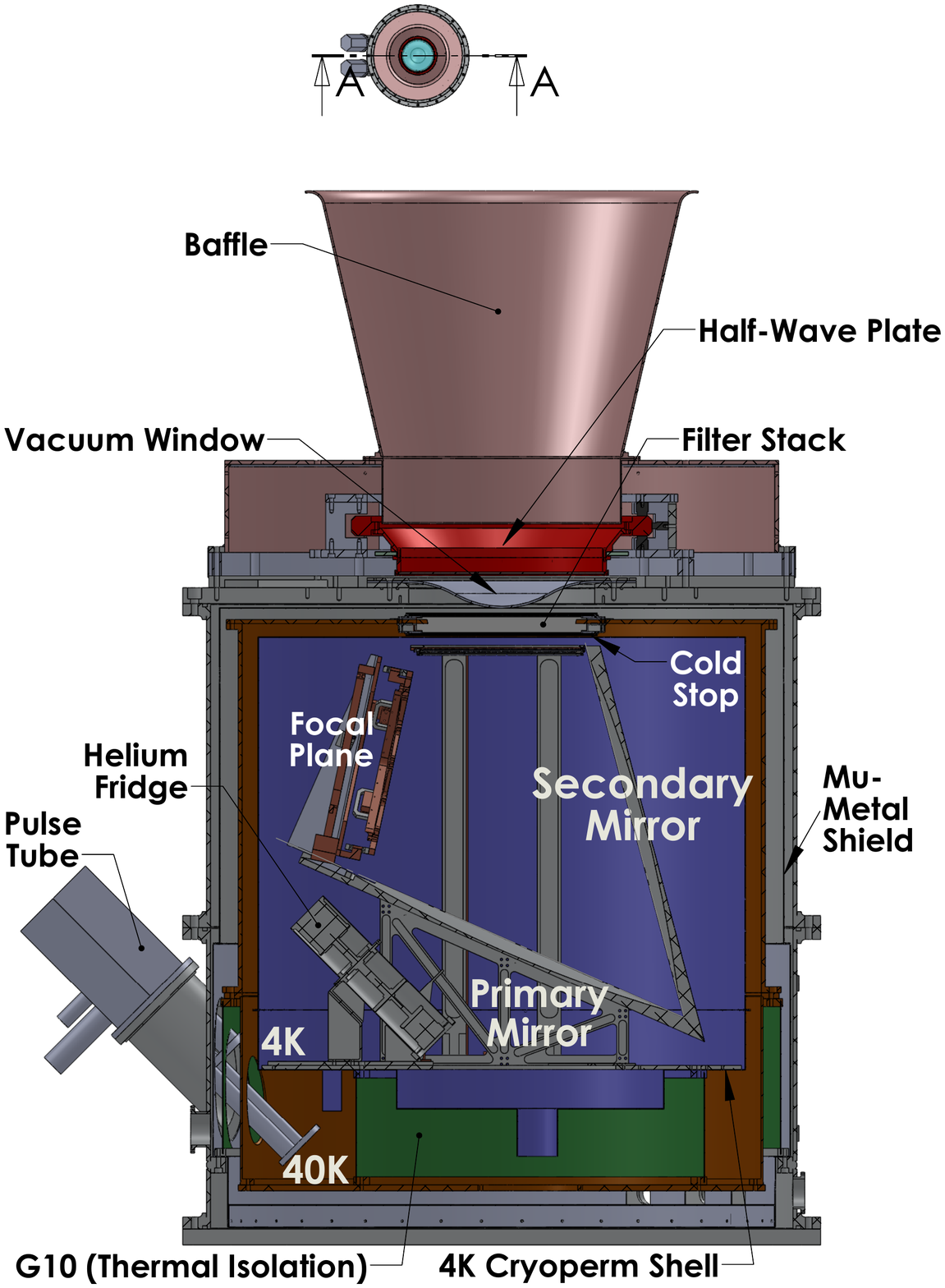}
		\caption{ABS receiver}
		\label{fig:receiver_overview}
	\end{subfigure}
	\hspace*{-2em}
	\begin{subfigure}[b]{0.6\textwidth}
		\centering
		\includegraphics[height=1\textwidth, height=0.55\textwidth,clip=true, trim=0in 0in 0in 0in]{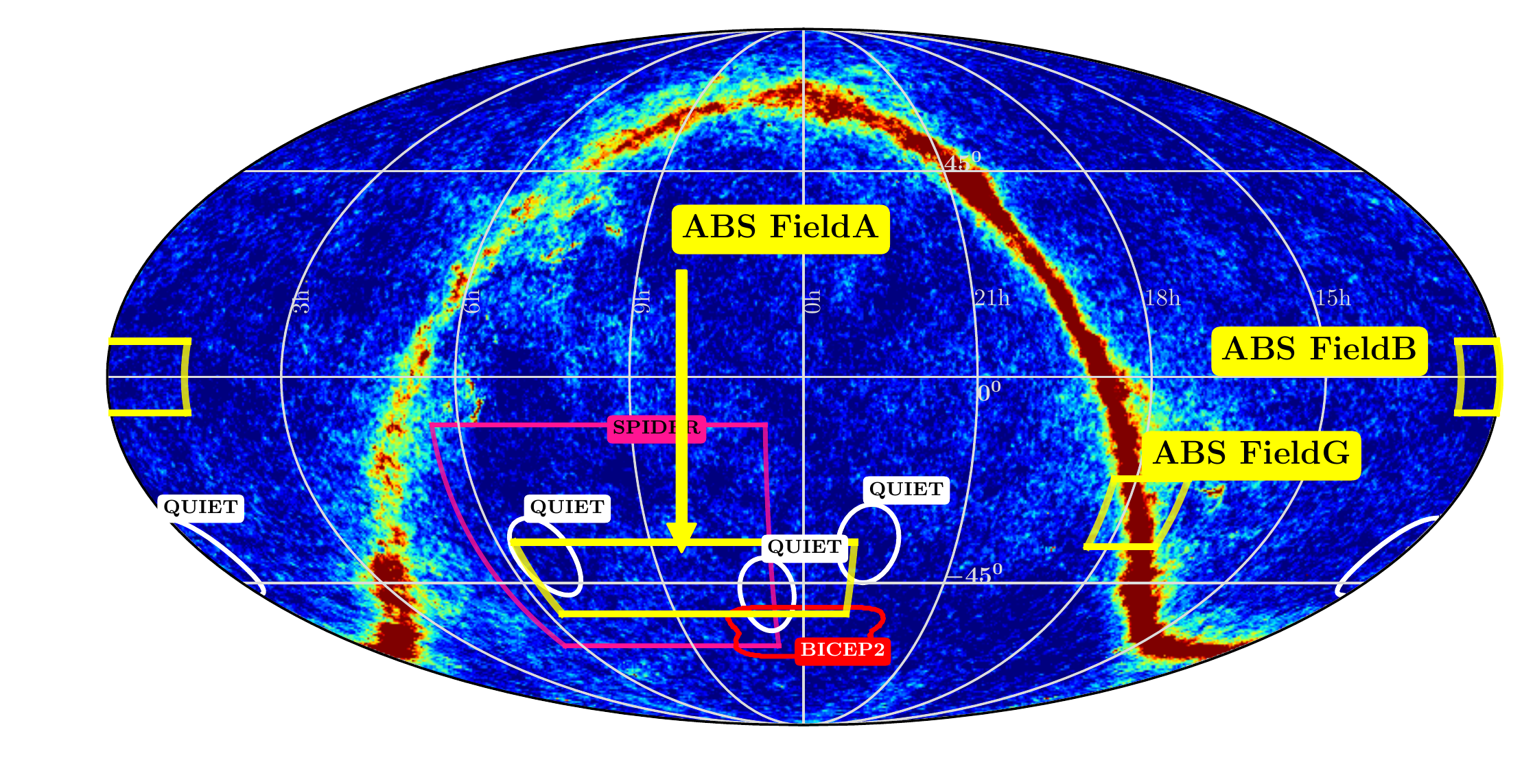}
		\caption{ABS observation fields}
		\centering
	\end{subfigure}
	\caption{On the left is an overview of the ABS receiver, showing key optical elements, including the primary and secondary mirrors, focal plane, cold stop, baffle and HWP. The nominal observing elevation of $45^{\circ}$ places the pulse tube cryocooler vertical. Magnetic shielding is provided by an ambient temperature mu-metal shield just inside the vacuum shell and a Cryoperm shell that doubles as the 4~K radiation shield of the cryostat. On the right panel are the ABS observation fields (in yellow) overlaid on the 143~GHz temperature map from Planck \cite{planck2015:hfipolmaps}, with color range from zero to 1~mK.   For context, the fields observed by several other small-aperture CMB instruments are also indicated: B2K \cite{jones:2006,bicep2:2014}; QUIET \cite{quiet:2011};  and SPIDER \cite{spider:2014}. See Table \ref{abs_obs_fields} for details.  This paper presents an analysis of ABS Field~A. }
	\label{fig:instrument_overview}
\end{figure}

The ABS instrument, shown in Fig.~\ref{fig:receiver_overview}, is a 145 GHz polarization-sensitive bolometric receiver and cryogenic telescope. Its key characteristics are presented in Table~\ref{abs_summary}.  It is integrated into a standard shipping container for rapid deployment. A hoist system elevates the  az-el mounted cryostat onto the roof of the container from where it scans the sky. A co-moving ground screen that shields the receiver from terrestrial radiation is attached after the mount is hoisted into position. The ground screen supplements a conical baffle at the window to the cryostat.  The cryostat is cooled by two pulse tube cryocoolers, and a  $^3$He/$^4$He system \cite{parker_thesis,penn_He3,lau_thesis} cools the detectors. A series of reflective metal-mesh and absorptive plastic filters block infrared radiation entering the cryostat window \cite{visnjic_thesis,essinger_thesis}.  The telescope optics comprise 60-cm primary and secondary reflectors maintained at  3.8~K  in a crossed-Dragone configuration~\cite{dragone:1978, dragone:1983}.  A 25-cm diameter aperture stop at 4~K terminates beam spill at a stable and cold surface. This configuration results in $32^{\prime}$~full-width-at-half-maximum (FWHM)  beams over a $22^{\circ}$ field of view. The focal plane of the receiver contains an array of 240 feedhorn-coupled pixels \cite{appel_thesis,yoon_truce,truce_omt,optical_truce,jappel_truce}, each with two transition-edge sensor (TES) bolometers (one for each orthogonal polarization) operating from a base temperature of 300~mK. The detectors were 
fabricated at National Institute of Standards and Technology (NIST). The array achieved a noise-equivalent temperature (NET)\footnote{The array NET is calculated from the median NET of each detector that passes the data selection; the number quoted here refers to observations of Field A.} of 40~$\mu {\mathrm K} \sqrt{\rm s}$ at a precipitable water vapor (PWV) of 0.5~mm.\footnote{The PWV is estimated with data from the Atacama Pathfinder EXperiment (APEX) Weather Monitor: \url{http://www.apex-telescope.org/weather/Historical\_weather/index.htm}.} The NET is referenced to the CMB blackbody temperature\footnote{All reported temperatures for the CMB and detector-related quantities are relative to the CMB. }  (2.725~K) and calculated from measured detector noise-equivalent powers, bandpasses, and  calibrated responsivities.  The NEQ (the sensitivity to a single Stokes parameter)  is 41~$\mu {\mathrm K} \sqrt{\rm s}$ due to the incomplete modulation of incident polarized signals.  (See \S\ref{sec:responsivity}.)  

The detectors are arranged in 24 triangular ``pods" with 10 feedhorns each.  Each corrugated aluminum feedhorn couples light onto an orthomode transducer (OMT) comprised of niobium probes suspended on a thin silicon nitride membrane. The OMT sends radiation from the two orthogonal linear polarizations along microstrip lines to separate TES bolometers. On-chip stub filters define the detector bandpass. The detector polarization angles were chosen to minimize cross-polarization from the optics and are predominantly at $\pm45^{\circ}$ to the horizon \cite{essinger_thesis}.  The top and bottom halves of the ABS array were fabricated in two separate batches (A and B, respectively), each with two fabrication wafers (1-4 and 1-11 for A, 1-14 and 1-15 for B). 
Due to an unexpected change in the transmission line dielectric constant between the two fabrication periods, detectors in batch B have bandpasses shifted up by $\sim$9~GHz relative to batch A. The detector bandpasses were measured with an {\it in situ} Fourier transform spectrometer. Further details on the bandpasses can be found in Table~\ref{abs_summary} and \cite{simon/etal:2014}\footnote{This paper updates the results in \cite{simon/etal:2014} by weighting the frequency response by the contribution to the final CMB spectrum and by adopting uncertainties consistent with this analysis.}. The bolometers are read out using superconducting quantum interference devices (SQUIDs) in a time-division multiplexing scheme \cite{NIST_TDM}. The array NET is dominated by 155 bolometers from batch A with a weighted per-detector average sensitivity of 580 $\mu {\mathrm K} \sqrt{\rm s}$. The remaining detectors contribute about a fifth of the total weight.  
\begin{table}
  \caption{Key Characteristics of ABS}
\small
\begin{center}
\renewcommand{\arraystretch}{1.2}
\begin{tabular}{p{4cm}c|p{4cm}c}
\hline
\hline
Avg. cent. freq. A(B) & 145~GHz (154~GHz) & Beam FWHM & $32^{\prime}$ \\
Bandwidth A (B) & 36~GHz (33~GHz) & Field of view & $22^{\circ}$  \\ 
Num of feeds/bolometers & 240/480 & HWP rotation frequency & 2.55~Hz \\ 
Array NEQ & 41 $\mu \text{K} \sqrt{\text{s}}$  & Polarization modulation & 10.2~Hz \\ 
Typ.  detector $f_{3dB}$ & 110~Hz & Longitude & 67$^\circ$47$^\prime$15$^{\prime\prime}$~W  \\ 
Azimuth scan speed &  $0.75^\circ$ per second & Latitude & 22$^\circ$57$^\prime$31$^{\prime\prime}$~S \\ 
Data sample rate & 199.36~Hz & Altitude & 5190~m \\ 
\hline
\end{tabular}
\end{center}
\label{abs_summary}
\end{table}

The HWP for ABS is made of single-crystal, $\alpha$-cut sapphire that is 330 mm in diameter, 3.15 mm thick, and anti-reflection coated with 305 $\mu$m of a glass-reinforced, ceramic-loaded PTFE composite (Rogers RT/duroid 6002). An air-bearing system allows the HWP to be rotated smoothly at 2.55~Hz, modulating the incident linear polarization on the detectors at  10.2~Hz. The angular position of the HWP is read out with $2.4^{\prime}$ resolution by a Gurley Precision Instruments glass-slide incremental encoder with an index mark for zeroing once per revolution. The HWP is the first optical element that light from the sky encounters on its way to the detectors, allowing for a clear separation of sky polarization from instrument polarization. The rapid modulation of the HWP results in stable timestreams of polarization data.  After demodulation, the detector data exhibit stability over 500~s time scales, corresponding to a median $1/f$ knee of  $2.0$~mHz \cite{kusaka:2014_HWP}.

The detector, housekeeping, and telescope position/pointing are synchronized to a single master clock. The HWP position is asynchronously sampled and interpolated to the master clock. More information about the ABS instrument may be found in \cite{ kusaka:2014_HWP, essinger-hileman/etal:2015, simon/etal:2014, Simon_LTD_2014, essinger_thesis, appel_thesis, visnjic_thesis, parker_thesis, simon_thesis, srini_thesis}.

\section{Observations}
\label{sec_observations} 
The ABS telescope is located at latitude 22$^\circ$57$^\prime$31$^{\prime\prime}$~S and longitude 67$^\circ$47$^\prime$15$^{\prime\prime}$~W at an elevation of 5190~m in Parque Astron\'omico Atacama in northern Chile. Between February 2012 and October 2014, ABS gathered CMB and calibration data for three seasons of duration 10, 11, and 6 months.\footnote{The first four months of the observations during the first season were performed using an absorbing entrance baffle, which introduced significant loading. The CMB observations from this period are not included in the analysis reported here. All remaining data were collected with a reflective entrance baffle.}  
We targeted three fields in the southern sky for CMB observations, and report on one of them here, as described below. 
We used the Moon, Jupiter, Venus, and the star forming \mbox{H\,{\sc II}} region RCW38 to derive the pointing model (\S\ref{sec:pointing}). 
Jupiter and Venus observations were also used for  beam characterization as discussed in \S\ref{sec:beams}.
The Crab Nebula, a polarized supernova remnant hereafter denoted Tau A, was observed to confirm the instrument's polarization 
properties (\S\ref{sec:detang}). 
Other calibration observations, which are described in more detail in \S\ref{sec:calibration}, fall into three categories: a) the use of a sparse wire grid for determining detector angles, responsivities, and time constants; b) detector current-vs-voltage (IV) curves for choosing the optimal biases for the detectors, tracking the loading and responsivity prior to each approximately hour-long observation, and obtaining the calibration constant for converting raw analog-to-digital converter (ADC) counts to power detected by the TES; and c) wide scans of the Moon for deriving the relative detector positions within the  array.  Sky dips, consisting of short-duration $5^\circ$ peak-to-peak scans in elevation at constant azimuth, were taken frequently, but not used for the analysis presented here.  ABS observes both during night and day.  

\label{sec:fields}
\label{sec:scan_strategy}
Table \ref{abs_obs_fields} lists the details of the ABS observation fields, while the right panel of Fig. \ref{fig:instrument_overview} displays them relative to other experiments.  To select the main and secondary CMB fields for observation, Field A and Field B, we identified low intensity regions 
in the dust maps from \cite{Finkbeiner1999} and optimized on field availability and survey uniformity.
When Fields A and B were unavailable, we observed Field G, a galactic patch to allow us to map polarized emission from the Milky Way, and a tertiary CMB Field C that has significant overlap with QUIET Field~4 \cite{quiet:2011}. 
In this paper, we report results from observations of Field A from our first two observation seasons. 

In the first season,  science observations took place between Sept 13, 2012 and Nov 20, 2012 (1634~h) and then again from Dec 28, 2012 through Jan 6, 2013 (209~h). In the second season ABS observed between March 29, 2013 and June 10, 2013 (1745~h) and then again from Aug 13, 2013 to Dec 21, 2013 (3135~h). Of this, the total time spent on Field A was 2398~h or 35\% of calendar time. 

\begin{table}
  \newcommand{\Tblh}{\rule[-1mm]{0mm}{5mm}}
\centering
\caption{Center positions, extents, areas and total observing times for the ABS fields for the first two seasons. Figures~\ref{fig:instrument_overview} and \ref{fig:maps} show the area covered.  Data selection is described in \S\ref{sec:data_selection}.}
\label{abs_obs_fields}
\begin {tabular} {crrrrrr}
\hline
\hline
\multirow{2}{*}{Field} & \multicolumn{1}{c}{RA ($\alpha$)} & \multicolumn{1}{c}{Dec ($\delta$)} & \multicolumn{1}{c}{$\Delta \, \alpha$} & \multicolumn{1}{c}{$\Delta \delta$} & \multicolumn{1}{c}{Area} & \multicolumn{1}{c}{$\Delta t$}\\
 & \multicolumn{1}{c}{[deg.]} & \multicolumn{1}{c}{[deg.]} & \multicolumn{1}{c}{[deg.]} & \multicolumn{1}{c}{[deg.]} & \multicolumn{1}{c}{[deg.$^2$]} & \multicolumn{1}{c}{[hours]}\\
\hline
\Tblh Field A &  25\hspace{1mm}\ & $-42$\hspace{1mm}\ & 90\hspace{1mm}\ & 25\hspace{1mm}\ & 2400\hspace{1mm}\ & 2398\hspace{1mm}\ \\
\Tblh Field B & 175\hspace{1mm}\ &     0\hspace{1mm}\ & 30\hspace{1mm}\ & 25\hspace{1mm}\ &  700\hspace{1mm}\ &  350\hspace{1mm}\ \\
\Tblh Field C & 341\hspace{1mm}\ & $-36$\hspace{1mm}\ & 25\hspace{1mm}\ & 10\hspace{1mm}\ &  250\hspace{1mm}\ &  390\hspace{1mm}\ \\
\Tblh Field G & 266\hspace{1mm}\ & $-29$\hspace{1mm}\ & 20\hspace{1mm}\ &  5\hspace{1mm}\ &   80\hspace{1mm}\ &  256\hspace{1mm}\ \\
\hline
\end {tabular}
\end{table}

Field A was given the top priority and was observed daily at an elevation of $\theta=45^\circ$ both as it rose in the east at azimuth $\phi=125^\circ$ and as it set in the west at $\phi=235^\circ$. While this strategy results in marginally cross-linked coverage, each map pixel is observed with almost a continuum of polarization orientations due to the continuously rotating HWP. The other CMB fields were observed when Field A was unavailable.  The $^3$He/$^4$He cryogenic cycle, which consisted of 36~hrs of observations followed by 7~hrs of recycling, was timed so that the recycling did not interfere with the $\sim$12~hrs/day of Field A observations.  Point source and other calibration observations sometimes superseded the observations of Fields B, C, and G. In addition, IV curves and sky dips were taken throughout the campaign.

The CMB data were collected with azimuthal scanning at constant elevation.  The scan speed was $0.75^\circ$ per second with a peak-to-peak scan amplitude of $10^\circ$ in azimuth ($7^\circ$ on the sky for elevation $45^\circ$). The HWP rotates $\sim$2.6 times before the telescope scans across one 
beam-FWHM-sized patch of sky. The scan center in azimuth is staggered by $4^\circ$  every other day.  At an elevation of $45^\circ$, this angle corresponds to the spacing between the centers of the 10-feedhorn pods ($3.4^\circ$ on the sky) minus half  the spacing between feedhorns in a pod ($1.25^\circ/2$ on the sky) at the center of the array. The staggering improves the uniformity of the sky coverage.

\section{Processing and selection of the time-ordered data}
We describe here the steps taken to convert the raw time-ordered data (or ``timestreams") into clean polarization data from which maps can be made.  
Since one of the key elements of ABS is the HWP, we begin with an overview of how the data are demodulated to isolate the polarization response.  Following that we next detail the data processing and selection.
The basic unit of processing is the constant elevation scan (CES);  the average CES duration is 70~mins, with a range of 20 to 100~mins. There are typically $\sim$150 back and forth scans in a CES.

\subsection{HWP modulation}
Because of the HWP modulation, the raw timestream from {\it each} TES detector measures the $Q$ and $U$ Stokes parameters as well as the intensity, $I$. When the ABS HWP rotates at $f=2.55$ Hz, the electric field direction 
is rotated at $2f$. Because the bolometer measures power, it produces the same signal for two oppositely directed electric fields and thus a polarized signal is modulated at $4f$. For a typical detector noise level, the atmospheric $1/f$ knee is near 1-2~Hz, well below $4f = 10.2$~Hz. The modulation eliminates the need for differencing the signals between pairs of detectors with orthogonal detector angles.

A HWP-modulated timestream $d_{m}$ can be described as
\begin{eqnarray}
\label{eq_hwp_data} 
d_{m}=I\  +\epsilon\ \mathrm{Re} \left[ m(\chi) (Q+iU) \right] +{\cal N}_{w}+ \ A(\chi), 
\end{eqnarray}
where $I$, $Q$, and $U$ represent the Stokes parameters of the incoming radiation,  $\chi$ is the angle between the frame of reference of the local polarization and the principal axis of the HWP,  $m(\chi) = \exp(-i4\chi)$ is the modulation function, 
$\epsilon$ is the modulation efficiency, and ${\cal N}_{w}$ is the detector white noise component. The last term, $A(\chi)$, encodes the angle-dependent emission, transmission and reflection properties of the HWP. The $A(\chi)$ predominantly consists of the $2 f$ component. Its amplitude depends on a combination of unpolarized sky, HWP, and receiver temperatures; it is affected by thermal drifts in the receiver.   As described in 
\S\ref{sec:responsivity}, the $2f$ component serves as a calibration tool for monitoring the detector responsivities.   Leakage of  $I$ into the $4f$ component of $A(\chi)$ has been constrained to be less than 0.07\% for the ``dipole" and ``quadrupole'' terms \cite{shimon/etal:2008}, corresponding to a systematic error contribution of $r < 0.01$, even without attempting to measure and remove the leakage signal~\cite{essinger-hileman/etal:2015}. 

\subsection{Processing steps and data selection}
\label{sec:data_selection}
The timestream processing is divided into four steps: 1) raw (intensity) timestream cleaning; 2) HWP demodulation; 3) $1/f$ noise modeling and filtering; and 4) removal of scan-synchronous structure.  We describe each below along with the associated data selection criteria. Detailed descriptions of those criteria are presented in \cite{simon_thesis,visnjic_thesis}.

The first step in the data selection process eliminates entire CESes if the telescope is not scanning properly, the HWP is not at its nominal rotation speed, there are no GPS data, or the detectors are not regulated at their nominal operating temperature. We eliminate all CESes for TESes that have zero responsivity over the entirety of both seasons. After these cuts, we have 708,725 TES-hrs  of data.  We next apply a series of data selection criteria that eliminate CESes for individual detectors (so-called TES-CES timestreams).  Finally, we only analyze CESes with more  than 150~TESes remaining.

\subsubsection{Raw timestream cleaning}

We make an initial pass at searching and repairing glitches in the timestream. Two-sample glitches, likely due to the readout electronics, appear as sharp spikes, while longer 10-sample glitches involve a jump over one or two samples followed by a thermal decay and are most likely caused by cosmic ray hits. These glitches, as well as the 0.1~s windows before and after the glitch periods, are flagged. These flagged samples are replaced with interpolated values for use in noise model estimation but are not mapped.\footnote{As described in \S\ref{sec:anal}, the power spectrum analysis relies on a noise model of ABS data. The effects due to masking are estimated in the simulation framework.}
The timestream is then searched for DC-level jumps from a failure of the flux-locked SQUID readout, which are similarly flagged and processed. The jump threshold is a level change of 100$\sigma$ or greater, where $\sigma$ is the median rms value of the difference between neighboring samples.

We then institute a series of cuts  that eliminate TES-CES timestreams if  they 1) do not have nominal IV curve characteristics, 2) have hundreds of glitch repairs, 3) have $>2$ SQUID jumps, 4) exhibit excess $1/f$ noise, 5) exhibit distributions that are not sufficiently Gaussian or stable (i.e.  high demodulated noise non-stationarity, raw noise non-stationarity, raw skewness, or raw kurtosis), or 6) do not have good time constant estimates derivable from the nearest IV curve (see \S\ref{sec:time_constants}).

Next, the detector samples are binned by the value of the HWP angle encoder over the full CES, producing an estimate of the $A(\chi)$ signal for each detector.  
The signal from the HWP at frequencies other than $4f$ serves as a continuous monitor of the health of the detectors~\cite{Simon_LTD_2016}.
For example, we find that the amplitudes of the $\sin (2f)$ and $\cos (2f)$ components of $A(\chi)$ are linear functions of the PWV during a CES. If they do not follow expectations, the detector is not biased and operating properly and the TES CES is cut.
 The $2f$ component is also used to recover the PWV in periods when the APEX radiometer was down and, as described in \S\ref{sec:responsivity}, to monitor the detector responsivities.   

The subtraction of $A(\chi)$ from the timestream is done with a truncated Fourier series in $\chi$ comprising the first 20 terms, including the mean.
At this stage in the processing, a filter is applied to deconvolve the  $f_{\mbox{\scriptsize 3dB}}$ = 60~Hz  antialiasing Butterworth filter in the data acquisition electronics \cite{appel_thesis, battistelli/etal:2008a,battistelli/etal:2008b}.   

\subsubsection{HWP demodulation}
\label{sec:demod_subsection}
Prior to demodulation, the timestreams are bandpass filtered around the \fourf polarization signal peak. The filter is a symmetric flat-top filter with cosine roll-off of width 0.1~Hz that turns on at $\pm$1.05~Hz around 4$f$. We optimized this filter to minimize loss of signal while avoiding unexplained narrow noise features in the Fourier domain near the signal band. 
To demodulate a timestream, we multiply $d_m$ by the complex conjugate of the modulation function, $\overline{m}(\chi)=\exp\left(i4 \chi\right)$. 
The bandpass-filtered timestreams are demodulated and then low-pass filtered at 1.1~Hz using a cosine roll-off complementary to the bandpass filter.  This second filtering step removes noise or spikes that could be introduced by the demodulation process. 
The filtered timestreams capture essentially all of the CMB signal since the scan speed is slow: $\sim$1~beam-FWHM/s.  Finally, the mean and slope of each demodulated timestream in each CES are removed. 
The $Q$ and $U$ Stokes parameters form the real and imaginary parts of the demodulated timestreams. Detailed descriptions of the demodulation technique and the HWP systematics are given in  \cite{kusaka:2014_HWP, essinger-hileman/etal:2015, essinger_birefringent_paper}. 
The cosmological results are based on the demodulated timestreams.

Each demodulated timestream is searched for periods of excess noise in the region of $f<0.2$\,Hz using short-time Fourier transforms applied to four-minute portions of the data for every two-minute step (thus they overlap in two-minute segments).  The same search is performed for ``$2f$-demodulated'' timestreams, those demodulated using $\exp\left(i2\chi\right)$ instead of 
$\exp\left(i4 \chi\right)$.   We then mask any four-minute segments in the $2f$- and $4f$-demodulated timestreams exceeding a selection cutoff for the noise level, resulting in 0.6\% of the total detector data being masked.  In practice, once the suspect region is identified, the mask is applied to the raw timestream prior to the demodulation. Masked portions are then filled with an interpolation estimate and not included in the maps. Timestreams where greater than 25\% of the data are masked due to noise flareups are treated as glitches and cut from further analysis.

Although $A(\chi)$ is approximately constant over a CES, it does slowly vary. This is accounted for as $1/f$ noise in the demodulated timestreams as described next.

\subsubsection{Noise modeling and filtering}  
\label{sec:noise_model}

The noise spectra of the demodulated timestreams are typically well-described by white noise plus a low-frequency noise term, which we model as 
$P_n(f) = A^2[1 + ( f_{\rm knee}/f )^n]$ with $f_{\rm knee}$ and $n$ determined for each CES TES.  Their typical values are $f_{\rm knee}=2$~mHz and $n=1.5$. To handle correlations in the $Q$ and $U$  timestreams of a single detector, we first diagonalize the 2-by-2 matrix formed from $Q$, $U$ auto- and cross-power spectral densities averaged in the band below 5~mHz, for each CES.  The diagonalization matrix can be thought of as performing a complex rotation of the polarization basis by an angle $\phi_o$. The rotated basis is given by  $Q' + i U' = \exp( i \phi_0) \left(Q + i U\right)$. The rotation angle is assumed to be constant as a function of frequency.

We next estimate the spectral densities of the primed timestreams. We fit the $1/\kern-0.1em f^n$+white noise model of the $Q^{\prime}$ and $U^{\prime}$ power spectra separately.  The rotation by $\phi_0$ maximizes the amplitude of $1/\kern-0.1em f^n$ noise in either the $Q^{\prime}$ or $U^{\prime}$ data. We find that nearly all of the temporally-correlated polarization noise exists in a single polarization mode of either the $Q^{\prime}$ or $U^{\prime}$. Thus we expect our assumption of a frequency-independent $\phi_o$ to be approximately valid. 
A frequency-domain inverse-variance weighting is applied to the 
Fourier transforms of $Q^{\prime}$ and $U^{\prime}$ using their respective noise power spectra fits. Then we apply the inverse of the $\phi_o$ rotation, followed by the inverse Fourier transform, and we recover inverse-variance weighted $Q$, $U$ timestreams to be used for mapping.

From the fitted noise model, we select CES-TES timestreams with low reduced $\chi^2$ for the noise fit both above and below 0.5~Hz (with the scanning frequency and its harmonics removed), with an additional requirement for selection that the fitted white noise and $f_{\rm knee}$ are nominal.  

\subsubsection{Scan-synchronous signal}   
\label{sec:SSS}
Finally, the $Q+ i U$ timestream is filtered by projecting out Legendre polynomial modes in azimuth over the full CES. The Legendre polynomial representation is used to estimate the scan-synchronous signal (SSS) from non-celestial sources. Possible contaminants to the data at the scan frequency include pickup of ground emission through diffraction, magnetic pickup, and a stable atmospheric signal that survives demodulation. The detected SSS, typically \lap$250\,\mu$K in amplitude, is projected out from the timestream by approximating it as a linear combination of the first 20 Legendre polynomials.

After subtracting the SSS model, CES TES timestreams with high reduced $\chi^2_{SSS}=\chi^2_{Real} +\chi^2_{Imag}$ compared to a residual model of zero are cut from further analysis. Entire CESes are also cut if the median $\chi^2_{SSS}$ of a subset of well-behaved detectors is too large. Additionally, CES TES timestreams with a large amount of slowly varying SSS, which is parameterized by the broad bump static SSS criterion, are cut. The broad bump statistic is defined as Fourier-domain excess in a range of $\sim \pm 12$\% of the scan frequency (37.5~mHz).   Further analysis showed that a fraction of data ($\sim 15$\%) might contain residual time-varying SSS on sub-CES timescales.  We show the negligible impact of the possible time-varying SSS in \S\ref{sec:systematics}. The cuts based on the SSS eliminate 7.1\% of the data that pass the previous cuts as defined in Table~\ref{tab:cut_stats}.

After calibration, the data processed as described above are coadded, and binned to form maps of $Q$ and $U$.
The processing operations described above lead to a reduced level of power in the resultant maps. Because of this, the transfer function of the map and the uncertainty on the power spectrum must be determined with simulations as discussed in \S\ref{sec:anal}.

\subsubsection{Data selection summary} 
Table~\ref{tab:cut_stats} summarizes the data selection criteria described above.  The order of the rows in the table gives the order in which the cuts are applied.    Overall, we begin with $\sim$ 2398~hrs of data on Field A, and nominally 479 TESes, for a total of  1,148,462 TES-hrs.  After all the data selection criteria are applied, there are 461,237 TES-hrs  remaining, which corresponds to 59.8\% of the total data being cut.  However, roughly half of that loss is due to the fact that only 351 of the TESes were functional.  

\begin {table}[b!]
\begin{center}
\caption{The impact of the successive application of each data-selection criterion for Field A. The first entry corresponds to 479 TESes observing for 2398 hrs. For each successive cut, the number of remaining CES-TES timestreams and the corresponding number of TES-hrs  after the cut are shown along with the percentage of the TES-hrs that have been cut from the previously retained data.} 
\label{tab:cut_stats}
 \begin{tabular}{ c  r  r  r }
    \hline\hline
    Cut Name  & \multicolumn{1}{c}{Number TES CESes\hspace{3mm}\ } & \multicolumn{1}{c}{TES-hrs\ \hspace{3mm}} & \multicolumn{1}{c}{$\%$ Cut} \\\hline
    (Total Number)                          &1,024,102\hspace{10mm}\  & 1,148,462 &    0$\%$\\
    Nominal Telescope Operation             &  908,184\hspace{10mm}\  & 1,059,268 &  7.8$\%$\\
    Non-Zero Responsivity                   &  666,348\hspace{10mm}\  &   777,208 & 26.6$\%$\\\hline
    Detectors Biased and Operating Properly &  606,650\hspace{10mm}\  &   708,725 &  8.8$\%$\\
    No Excess Glitches                      &  520,711\hspace{10mm}\  &   606,916 & 14.4$\%$\\
    Nominal SSS                             &  483,748\hspace{10mm}\  &   564,098 &  7.1$\%$\\
    Gaussian and Stable                     &  446,676\hspace{10mm}\  &   520,169 &  7.8$\%$\\
    Nominal White Noise Properties          &  418,763\hspace{10mm}\  &   487,277 &  6.3$\%$\\
    Detectors Not Under Excess Loading      &  407,912\hspace{10mm}\  &   475,094 &  2.5$\%$\\
    Cut CES if $<$ 150 timestreams          &  396,047\hspace{10mm}\  &   461,237 &  2.9$\%$\\\hline
     \end{tabular}
\end{center}
\end{table}

\section{Characterization and calibration}\label{sec:calibration}

In this section, we begin with a discussion of the pointing solution, describe the beam profiles and window function, and discuss how we determine the detector responsivities, time constants, and polarization angles. Systematic errors associated with these are identified and tested with alternate models, as described in \S\ref{sec:systematics}.  

An important calibration tool for ABS was a sparse wire grid \cite{tajima/etal:2012, quiet:2011, kusaka:2014_HWP}, occasionally inserted above the HWP or above the baffle, depending on the purpose. The grid serves multiple purposes as discussed in \S\ref{sec:responsivity},
\S\ref{sec:time_constants}, and \S\ref{sec:detang}. 
The grid is composed of thin, reflective, parallel manganin wires spaced at intervals of 2.5~cm.
It is a source of polarized radiation with a direction that can be rotated around the line of sight.
The emission from the wire grid can be complex, but we found empirically that the signal was polarized in the direction parallel to the wires.
Hence we conclude that the signal is the power emitted by the $\sim270$~K wires plus power emitted from below the wire grid that is reflected back into the cryostat. The beams uniformly illuminate a $\sim$25~cm diameter circle near the center of the wire grid (a similar footprint to the aperture stop) sampling roughly 10 grid wires.   

\subsection{Pointing} \label{sec:pointing}
The boresight pointing solution was obtained using $\sim$150 observations of the Moon, Jupiter, RCW38, and Venus made at different positions on the sky. 
The data were fit either in the time domain (Moon, Jupiter, and Venus) or in the map space (RCW38),  depending on the signal-to-noise ratio (SNR) of the observation. For RCW38, the maps from multiple detectors during a single observation were stacked to improve the SNR. The centroids of the fits were compared to the ephemerides 
of the sources to determine the azimuth and elevation offsets for each observation. A six-parameter pointing model was fit, 
characterizing the physical imperfections of the telescope. 
The residual boresight pointing uncertainty is $0.04^\circ$, less than a tenth the beam FWHM.  
Assuming a Gaussian distribution, we convolve this pointing uncertainty with the beam to obtain the net beam window function. The relative pointing among detectors was determined with seven 
$\sim$40$^\circ$ wide, $\sim$2~hr constant-elevation scans of the Moon; the relative pointing uncertainty is $0.01^\circ$. The boresight pointing was continuously monitored for seasonal variations. There were three instances of unexplained encoder zero point shifts, which were identified and corrected almost immediately using Moon and Jupiter observations. A small fraction ($\sim6$\%) of the associated data were cut due to ambiguous pointing. 
No other significant seasonal variations were detected. More details on the pointing model may be found in \cite{srini_thesis, simon/etal:2014}.

\subsection{Beam profiles and window functions}
\label{sec:beams}
The beam profiles and window functions are determined through a combination of modeling and observations of Jupiter. Because of ABS's limited elevation range, Jupiter is routinely visible only in the upper half of the array where the most sensitive detectors are located. Observations of Venus, which include the lower half of the array, are consistent but do not have sufficient statistical weight to inform the profiles. Although the electric field pattern in the cryostat is determined by the feedhorns and the reflectors, the far-field beam profiles are largely determined by truncation on the 25 cm diameter absorbing aperture stop at 4~K. The aperture stop, with edge taper between $-7.5~{\rm and}~-10.5$~dB depending on the feedhorn location, is between the ambient temperature HWP and the cold primary mirror. (See Fig.~\ref{fig:receiver_overview}.)  

\begin{figure}[t!]
\centering
\includegraphics[width=0.7\textwidth]{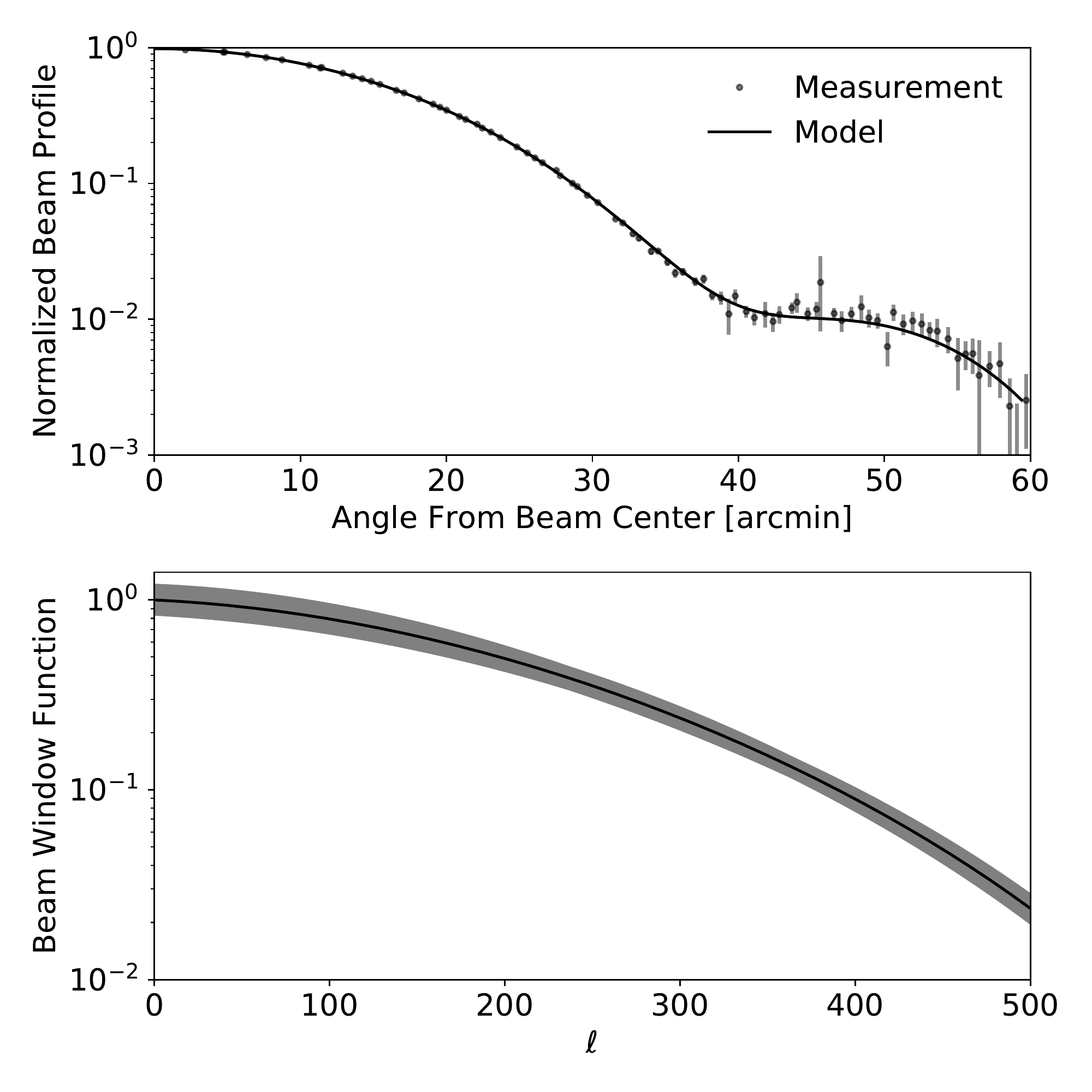}
\caption{{\it Top:} The beam profile from Jupiter observations with detectors near the center of the focal plane. The data are overplotted with a fit to the beam model described in the text. The forward gain is $50$\,dBi. {\it Bottom:} The beam window function with uncertainties shown by the gray band. The calibration is done with Jupiter, which determines the product of the detector responsivity and the forward gain to 6\% accuracy 
(see \S\ref{sec:responsivity}). The combined beam and calibration uncertainty in the power spectrum is 15\% at low $\ell$, reaches a minimum 
of 12\% near $\ell=350$, and then increases at larger $\ell$. The uncertainty at $\ell=350$ is dominated by that of the amplitude in Jupiter measurement.  The $\ell$ dependence of the uncertainty is driven by the uncertainty in the beam calibration.}
\label{fig:beam}
\end{figure}

We use results from a physical optics code (DADRA \cite{YRS:DADRA}) and  a ray-tracing code (Zemax\footnote{Zemax, LLC; Kirkland, WA 98033;  \url{http://www.zemax.com}.})  to develop a simple two-parameter model of the fields in the aperture for the feedhorns in each pod. This model accounts for the leading terms in the intensity profile and phase aberration, which effectively parametrize the width and sidelobe level of the azimuthally-symmetrized far-field beams. We then combine the measured and modeled profiles, weighted by the contribution that each pod makes to the map, to get an overall effective beam profile. The detectors for the feedhorns in the lower half of the array (from batch B) contribute only 20\% of the total statistical weight.  A Gaussian profile with $\theta_{FWHM}=32.1^\prime\pm0.4^\prime$ is a good approximation to the main beam and the effective average solid angle is $\Omega_B=101\pm3\mu$sr.   
The uncertainty on $\Omega_B$ is dominated by an estimate of the systematic error in the measurements plus model.
This estimate updates the presentation in \cite{simon/etal:2014}. The beam profile and window function are shown in Figure~\ref{fig:beam}.

Pickup through the sidelobes was controlled with a conical reflective baffle (Figure~\ref{fig:receiver_overview}) and a rectangular co-moving ground screen 1.5~m$\times$1.2~m by 1.2~m tall. To enter the aperture from the ground, direct rays have to diffract twice, once over the ground screen and then once over the baffle. Based on approximating the ground screen as a knife edge and applying Sommerfeld's solution to radiation from the ground, we estimate an equivalent of 40 mK (Rayleigh Jeans) of radiation incident on the aperture if the baffle were not in place and not accounting for the beam gain. Diffracted radiation from the ground is primarily polarized in the vertical direction\footnote{The ground screen also slightly polarizes the diffracted emission from the atmosphere, but at a lower level, and it reflects the atmosphere into the sidelobes and polarizes it.}. An accurate accounting of the diffraction from the baffle is involved as all the relevant edges are in the near field.  Neglecting near-field effects and without the baffle in place, calculations show that the total diffracted power from the ground into the aperture, based on the far-field beam profile,  would be $\sim 650~\mu$K and $\sim60~\mu$K for the vertical and horizontal polarizations respectively. Variations in the ground emission temperature, which we estimate at 5\% or 15~K, would enter as a vertically polarized signal at the $\sim35~\mu$K level.\footnote{To convert to a Rayleigh-Jeans temperature divide by 1.64.} The baffle reduces this further.

It is possible that there are paths from the ground into the detectors that we have not considered. We did not carry out {\it in-situ} measurements and so we cannot rule them out.  The timestream filter we apply projects out such ground-fixed signal (\S\ref{sec:SSS}).  Residual systematics after this filter would be a repeatable polarized signal over 10$^\circ$ azimuth scans centered on azimuths $125^\circ$ and $235^\circ$ imprinted throughout the maps.  We saw no evidence of such systematics. In addition, the data passes a number of null tests (e.g., day vs. night, east vs. west, \S\ref{sec:nulltests}) that we would expect to fail if there were measurable contamination through the sidelobes.

\subsection{Responsivity}\label{sec:responsivity}

The responsivity is determined in six steps: 1) analysis of $\sim$hourly IV curves for preliminary calibration of the timestreams into  power (aW); 2) tracking of the time variability of the responsivity with the $2f$ signal; 3) approximately monthly observations of Jupiter to provide absolute calibration 4) determination of the relative detector responses to a Rayleigh-Jeans signal provided by occasional sparse wire grid measurements (flat fielding); 5) conversion to CMB temperature units using the measured bandpasses for detectors from each fabrication wafer; and 6) an analytic estimate of the efficiency of the HWP. The net calibration uncertainty is 6\% in Stokes $Q$, $U$.  We confirm the calibration through 
cross-correlation with Planck as described in \S\ref{sec:cfPlanck}. 

The responsivity may vary as a function of time because of the variations in atmospheric loading, the cryogenic state, or aging of the detectors over the course of observing seasons. The IV curve calibrations correct for slow ($\sim$hour timescale) changes in the atmospheric loading.  We examine remaining time-dependent variations in two ways:  using the $2f$ signal (our primary method) and with a secondary bias power fit (BPF) method. A supplemental third method using the sparse-wiregrid calibration, which examines the time dependence of only the relative responsivity, is also used for consistency check. The $2f$ method uses the amplitude of the $2f$ signal in a given epoch to 
track the time variability of the responsivity. The $2f$ signal comes from a combination of differential emission, differential transmission, and 
differential reflection for the ordinary and extraordinary axes of the HWP. This signal can be broken up into a component that depends on total sky loading, determined by the PWV, and one which does not. We fit the data with a linear function of the PWV and empirically determine the dependence in each epoch. We track the fitted $2f$ intercepts from the amplitude versus sky loading curves within each epoch and assume 
that the model is unchanged over the course of the epoch. We thus attribute variations in the $2f$ intercepts between epochs to variations in the 
responsivity~\cite{Simon_LTD_2016}.

The BPF method relies on bolometer power equilibrium: $P_{\rm therm} = P_\gamma + P_{\rm bias}$. Thus, for a fixed flow of power, 
$P_{\rm therm}$, to the thermal bath, changes in the loading $P_\gamma$ must be compensated by changes in the electrical bias power $P_{\rm bias}$ applied to the TES. For each detector, we perform linear fits of $P_{\rm bias}$ from the IV curves versus PWV and use the slopes to trace time variability in detector responsivity.  For both methods, we only fit data with PWV $< 2.5$~mm to avoid nonlinear detector responses that can occur under high loading. We normalize the tracers (BPF slopes and $2f$ intercepts) by their values in a reference epoch as described next. 

Over the first two seasons, the batch-B detector responsivities decayed in four discrete shifts. Although we have not yet identified the source of the variability, the shifts each occurred after a period of a month or longer in which no observations were made and the cryostat warmed to ambient temperature. The shifts were seen in the $2f$ response, IV curves via the BPF, and wire grid measurements. We enumerate the four epochs of stable responsivity from $n=1$ to $n=4$, and identify $n=4$ as the reference epoch, as it includes the most (five) wire grid measurements.
The median percentage shifts between $n=1\rightarrow 2$,  $n=2\rightarrow 3$, $n=3\rightarrow 4$  are $-28$\%, $-38$\%, and $-36$\% respectively for batch-B detectors whereas the batch-A  detector are consistent with constant responsivity.

During the campaign, we performed 10 measurements of the relative detector responsivities (and of the polarization angles) by inserting a sparse wire grid in front of the HWP and baffle and rotating it in discrete steps. The demodulated ($Q$ and $U$) time streams of the detectors respond sinusoidally at twice the grid rotation frequency. The relative responsivity 
$r^{wg}_{i,n}$ is determined by the ratio of the amplitude of detector $i$'s response to that from the reference detector ($d_{\rm ref}$).  The ratio for each detector is averaged  over the wire grid calibration measurements within each epoch $n$; its uncertainty  is estimated from the spread in multiple measurements. These uncertainties are treated as Gaussian and propagated into the final power spectrum systematic errors in \S\ref{sec:systematics}.

We model the absolute responsivity $R_{i,n} $ of detector $i$ for  epoch $n$ as
\begin{equation}
R_{i, n} = R^{RJ}_{d_{\rm ref}} \, r^{wg}_{i,4} \left ( \frac{ b_{i,n} } { b_{i,4} } \right ),
\end{equation}
where $b_{i,n}$ is the responsivity for detector $i$ in epoch $n$ obtained from the $2f$ method and 
$R^{RJ}_{d_{\rm ref}} =126 \pm 6.3$~aW/mK as determined with Jupiter. For Jupiter we use a brightness temperature of $T_J = 173.6\pm0.92$~K and solid angle $\Omega_J=2.481\times10^{-8}(5.2/d)^2$, where $d$ is the distance to Jupiter  in Astronomical Units (A.U.) at the time of the measurement \cite{planck_planets:2016,weiland/etal:2011}.  The uncertainty on Jupiter's temperature is negligible compared to our overall calibration uncertainty.  The model using the $2f$ method is consistent with the BPF method.
 
The HWP does not modulate linear polarization with 100\% efficiency.  (In equation~\ref{eq_hwp_data}, $\epsilon < 1$.)   This creates a small difference in responsivity for polarized and unpolarized sources. The HWP polarization modulation efficiency is computed to be 97\% for A-batch detectors and 92\% for B-batch detectors from a transfer-matrix model~\cite{essinger-hileman/etal:2015}.  An effective polarization efficiency, which is an average accounting for the data selection efficiency and the inverse variance weighting of the detectors, is 96.5\%.

Conversion from a calibration based on Jupiter's brightness temperature to a CMB-referenced temperature depends on the frequency of the observations. For the batch-A detectors, $\nu_{cent}=145\pm1.1\,$GHz with a corresponding $\delta T_{\rm CMB}/\delta T_{\rm RJ}=1.64\pm0.02$. This leads to  a 3.1\% uncertainty in the conversion factor. The batch-B detectors contribute much less weight, and so do not increase the overall uncertainty.  We combine errors in quadrature: 5\% for the absolute calibration, and 3\% each for the errors in the central frequencies and in the beam solid angle. The combined calibration error relative to CMB temperature fluctuations has a net 6\% uncertainty at $\ell=350$, corresponding to a 12\% uncertainty in the CMB power spectrum (Figure~\ref{fig:beam}). In  \S\ref{sec:systematics} we assess possible systematic effects associated with the time dependence of the responsivity by comparing to a constant responsivity model. 

\subsection{Time Constants} \label{sec:time_constants}
The phase $\psi$, which relates a raw demodulated timestream to an angle-calibrated $Q+iU$ timestream, is given by the time delay of the detector response and a constant offset $\psi_0$ that is related to the detector polarization angle. It can be modeled as the phase of a one-pole filter:
\begin{eqnarray}\label{eqn:tc}
  \psi &=&\psi_0+\arctan{(4f/f_{3\rm dB})} \\
  &\approx&  \psi_0 + \frac{4f}{f_{3\rm dB}}\qquad {\rm if} \,\, (4f)^2 \ll f_{3\rm dB}^2\,,
\end{eqnarray}
where the 3dB frequency $f_{3\rm dB}$ is inversely related to the optical time constant $\tau_{opt}$ by $2\pi f_{3\rm dB}=1/\tau_{opt}$~\cite{Simon_LTD_2014}. Because ABS uses a HWP to modulate the polarization, fluctuations in the detector time constants due to varying atmospheric loading cause phase shifts in the polarization signal, which result in shifts in the measured detector polarization angles. If $f_{3\rm dB}$ is small, the time constants must be accounted for in the polarization angle calibrations and in the signal demodulation at 10.2~Hz.

By slowly varying the rotation speed of the HWP with a sparse wire grid in place to input a polarized signal, we made accurate {\it in situ} optical $f_{3\rm dB}$ measurements using the phase lag of the $4f$ signal. 
The IV curves can be used to estimate  the instantaneous proportionality constant  $\eta^{-1}$ between $\tau_{opt}$ and $\tau =C/G$, the intrinsic bolometer time constant, defined in terms of its thermal capacitance $C$ and conductance $G$.
 This method is first used to translate wire grid measurements of $\tau_{opt}$ into estimates of $\tau$ (which should be independent of time and loading).  Each detector's $\tau$ is combined with an estimate of  $\eta^{-1}$ from the IV curve before each CES to recover $\tau_{opt}$~\cite{simon_thesis}. The median $f_{3\rm dB}$ of the detectors is 109~Hz with 95\% above 55~Hz. For a detector with $f_{3\rm dB}=30$~Hz, a 10\% shift to a lower $f_{3\rm dB}$ would result in only a $0.96^{\circ}$ shift in the polarization angle of the detector.   Timestreams with $f_{3\rm dB}<30$~Hz are eliminated in the data selection, as are those where estimates of $\eta^{-1}$ are unphysical.

For our scan rate on the sky, we find that $\ell\approx 680\,f$; 50~Hz is at $\ell\approx 34,000$, well outside of the beam roll-off.  (See Fig.~\ref{fig:beam}.)
Thus, impacts other than that on the polarization angle variation, such as changes in the window function or pointing due to time delay,
are negligible.

\begin{figure}[!ht]
\centering
\includegraphics[width=0.8\textwidth, clip=true, trim=0in 0in 0in 0in ]{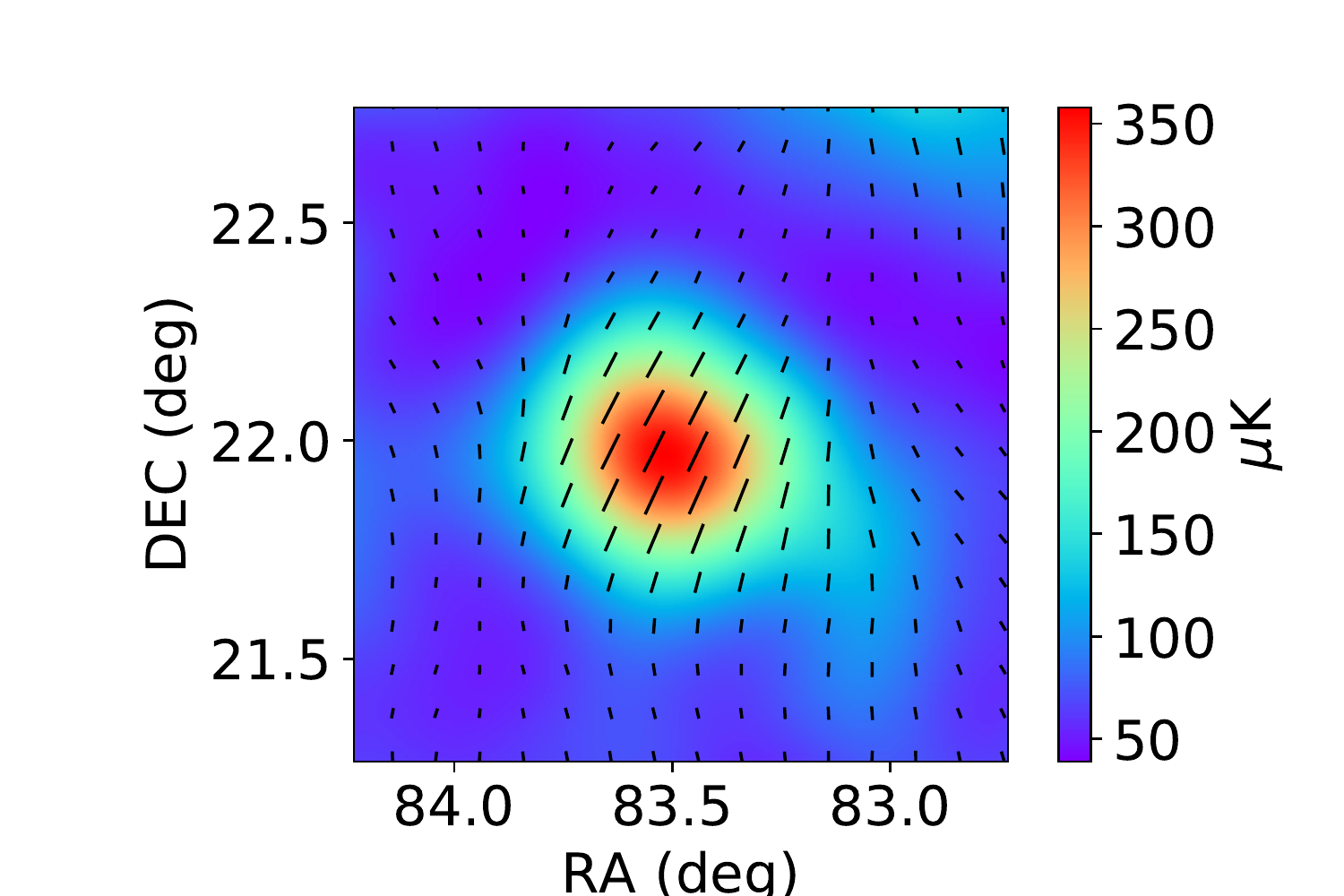}
\caption{ Map of Tau A. The color scale shows the magnitude of the polarization in CMB temperature units. The peak polarized amplitude at the ABS resolution is $308\pm14({\rm stat})\pm22({\rm sys})~\mu$K relative to the CMB. We measure the polarization direction to be 
$\gamma_p=150.7^{\circ} \pm 1.4^{\circ}({\rm stat})$ determined from the best fit $Q$ and $U$ profiles.}
\label{fig:tauA}
\end{figure}

\subsection{Detector angles}
\label{sec:detang}
We measure the detector polarization angles using the sparse wire grid. Throughout the observing campaign, in addition to the ten rotating grid measurements we performed seven aligned wire grid calibrations.  The data from the rotating grid viewed through the HWP yields  the relative detector angles from the phases of the sinusoidal polarization signals and the relative responsivities from their amplitudes (as described in \S\ref{sec:responsivity}). For the aligned grid calibrations, the grid is aligned to reference marks on the cryostat that are tied to the direction of gravity, with an uncertainty of $0.9^\circ$.  Using this wire grid reference, we constrain the absolute detector angle to $\pm1.1^\circ$. The measured absolute angle is consistent throughout the seven runs to within errors. Using Equation~\ref{eqn:tc}, the detector angle estimates are corrected for time constants~\cite{simon_thesis}. 

Tau A is the strongest polarized celestial source available to ABS. Approximately once per week during the first two seasons it was observed with the reference detector pair and its close neighbors. Using the detector angles from the wire grid we obtained the map of Tau A shown in Fig.~\ref{fig:tauA}. The timestream processing for map making here is mostly the same as the CMB analysis, except that a wider bandwidth of 2\,Hz is used for bandpass and lowpass filters in the demodulation process (see \S \ref{sec:demod_subsection}).   This larger bandwidth is introduced in order not to skew the structure in the map, or the shape of Tau~A.  With the 2\,Hz bandwidth, the scan speed, and the beam size, we expect the filters to have negligible impact on the measurements described in the following.
We fit Gaussian profiles with widths fixed at $\theta_{FWHM}=0.53^\circ$ to maps of $Q$ and $U$ to find $Q=160 \pm 15 \pm 11~\mu$K and 
$U= 263\pm 14\pm 18~\mu$K where the first error is statistical and the second is from the 6\% calibration at these angular scales. The reduced $\chi^2$ with 238 degrees of freedom is 0.77 for $Q$ and 0.94 for $U$. Computing the angle directly from these fits we find $\gamma_p=150.7^{\circ} \pm 1.4^{\circ}({\rm stat})$.
With a uniform weighting inside a radius equal to the $\theta_{FWHM}$  we find 
 $\gamma_p=152.7^{\circ} \pm 1.9^{\circ}({\rm stat})$.
This result is consistent with other observations at 150~GHz within the quoted errors~\cite{IRAM/TauA, macias-perez/etal:2010, actpol:2014, weiland/etal:2011, polarbear:2014}.  In particular, if one smears the Planck 143~GHz maps to the ABS resolution, pixelizes to $N_{\rm side}=512$ \cite{gorski/etal:2005} to match the resolution of the Tau A maps, one obtains an amplitude in P of $332~\mu$K.

For many early-universe models, though not all, $C^{EB}_{\ell}$ is expected to be zero.  The rotation of the polarization angle required to null $C^{EB}_\ell$ for a set of CMB polarization maps can be computed.   After unblinding the analysis (see \S\ref{sec:anal}), we find that an angular rotation of $-1.7^\circ \pm 1.6^\circ$ enforces $C^{EB}_{\ell}=0$. As this deviation is not statistically significant we do not correct for it.

\section{Analysis and Results}
\label{sec:anal}

We used a blind analysis strategy for ABS.  We developed and validated the pipeline for producing the ABS power spectra  without examining the TB, EB or BB spectra.  Validation included a series of null tests and systematics tests.   After finalizing the ABS spectra, we cross-correlated ABS E-mode data with Planck 100 and 143 GHz data on the same region of sky.  The Planck noise level is somewhat lower than that achieved by ABS and so this serves as an independent check of the ABS analysis. 
We also compared with the WMAP 23 GHz and Planck 353 GHz data to estimate the level of foreground contamination. Lastly, we unblinded the ABS TB, EB, and BB spectra, checked the overall polarization angle, compared to Planck B-mode data, and extracted an upper limit on $r$ from the ABS B-modes.

While the Planck maps for polarization are public, they urge caution in over interpreting the 100, 143, and 217 GHz maps for scales $\sim10^\circ$ and larger \cite{planck_overview:2015}. We interpret their caution to mean our comparisons should be considered preliminary. 

 We describe the pipeline and give the power spectrum results in \S\ref{sec:tod2ps}.  Null tests and systematics tests are described in \S\ref{sec:nulltests} and \S\ref{sec:systematics}, with a summary plot of the systematic errors in Fig.~\ref{fig:syst}. The comparison to Planck CMB data is found in \S\ref{sec:cfPlanck};   the foreground estimation is in \S\ref{sec:foregrounds}; and \S\ref{sec:lcdm} assesses the consistency of the ABS EE spectrum with the Planck $\Lambda$- cold-dark-matter (LCDM) model \cite{planck_overview:2016}.   Finally, the ABS limit on $r$ is presented in \S\ref{sec:r_limit}. 
 
\subsection{ABS timestream to spectra}
\label{sec:tod2ps}
\begin{figure}[t!]
\centering
\includegraphics[width=1.0\textwidth]{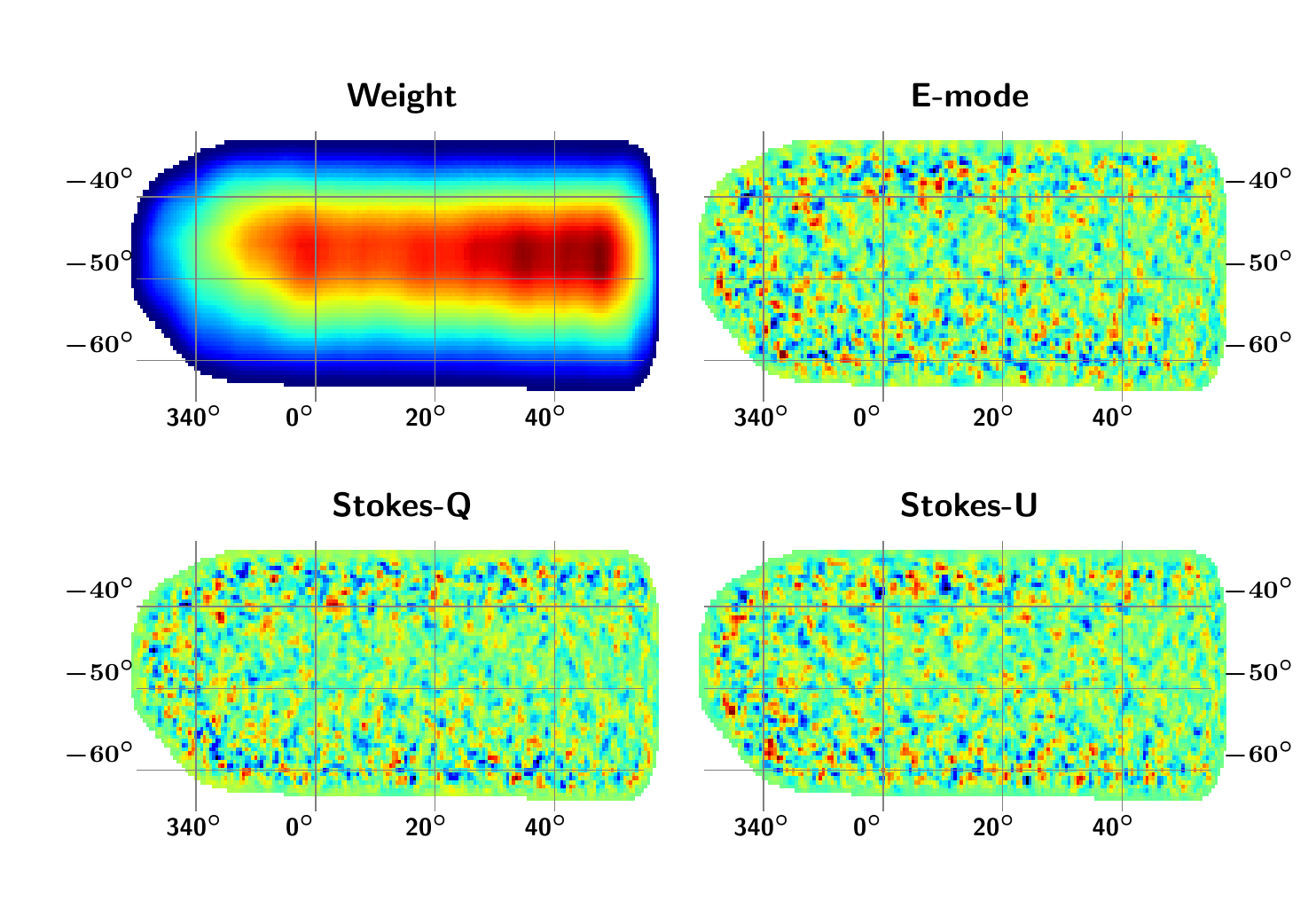}
\caption{The ABS maps for Field A in equatorial coordinates. {\it Top:} Normalized hit count ({\it left}) with a map of the E modes ({\it right}). 
{\it Lower left/right:} Stokes $Q$/$U$ maps. For the CMB maps, an isotropic high pass filter that passes  $\ell>30$ is applied. The color scale, blue to red, spans $-4~\mu$K to $4~\mu$K.}
\label{fig:maps}
\end{figure}

The demodulated time-ordered data that pass the cuts (\S\ref{sec:data_selection}) are projected onto the sky using the pointing model 
(\S\ref{sec:pointing}).  The per-pixel value returned by the mapmaker is the inverse variance-weighted average of pixel samples. This variance corresponds to the white noise level in the demodulated timestream, which is equal for the $Q$ and $U$ components. Because the maps are made directly from binning the processed data, comparing to them requires ``reobserving" any external data set, for example a WMAP or Planck map, using the ABS analysis pipeline. We use HEALPix pixelization in equatorial coordinates with $N_{\rm side}$ = 256 \cite{gorski/etal:2005}. The maps are shown in Figure~\ref{fig:maps}.

We use the MASTER algorithm \cite{Hivon2002}, as modified for polarization\cite{brown/etal:2005}, for estimating the true C$_\ell$ from pseudo-C$_\ell$ spectra.
Its basis is the Monte Carlo (MC) simulation of pseudo-C$_\ell$ spectra taking into account data processing and the ABS noise model. The pipeline allows for fast estimation of the experimental transfer function. 
In the MASTER algorithm, the power spectrum $\tilde{C}_\ell$ of the map with coefficients $\tilde{a}_{lm}$ is related to the true sky power spectrum C$_{\ell}$ by:
\begin{equation}
\bigl<\tilde{C}_\ell\bigr> = \sum_{\ell'} M_{\ell\ell'} F_{\ell'} B^2_{\ell'} \bigl<C_{\ell'}\bigr> 
\label{eq:61}
\end{equation}
where $M_{\ell\ell'}$  describes mode-mode coupling due to the geometry of Field~A and the weighting of the pixels within it, $F_\ell$  is the signal transfer function accounting for all timestream-level filtering described in \S4.2 and this section, and $B_\ell$ is the harmonic-space window function incorporating the ABS beam geometry and pixelization effects. The ${C}_\ell$ estimator in MASTER is derived from inverting the factors multiplying $C_{\ell}$  in the above equation. The $\tilde{C}_\ell$ and conversion factors to the $C_{\ell}$ estimator are binned to produce a binned power spectrum estimate, $\hat{C}_b$ where $b$ is the $\ell$-space bin number. There is no additive noise bias included in Equation~\ref{eq:61} because the power spectrum estimates are derived from cross-correlating sky maps formed from three-day intervals of observation. Before computing the spectra, a point source mask was applied as discussed in more detail below. The full set of 82 maps is then combined into a cross-correlation estimator. There is a small increase in error bars from ignoring autocorrelation terms.

\begin{figure}[t!]
\centering
\includegraphics[width=\textwidth]{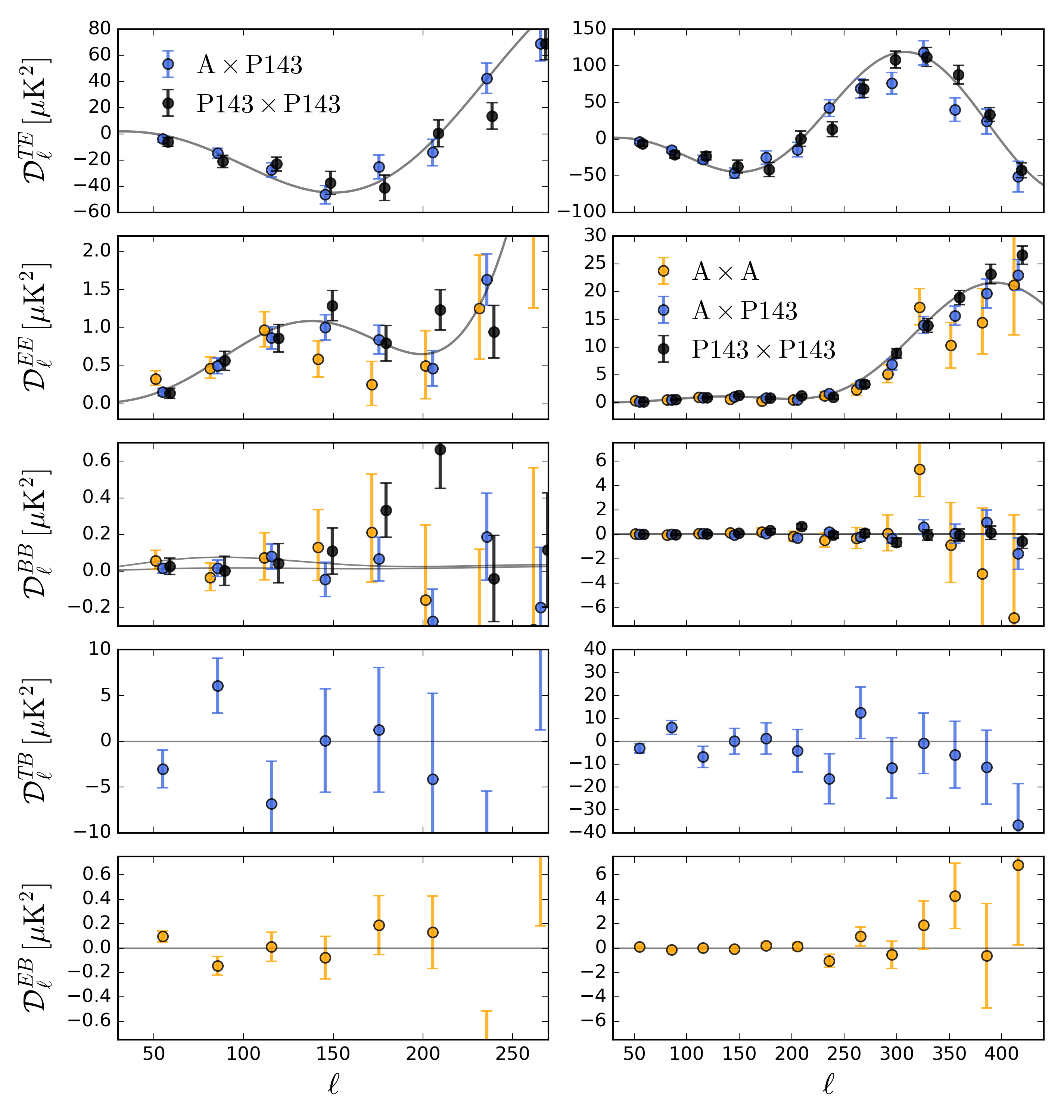}
\caption{The ABS and ABS$\times$Planck power spectra. For Planck we show spectra based on the reobserved Planck maps. The solid lines are for the LCDM model. The top curve in the BB panel is for $r=1$ plus lensing; the bottom curve is for $r=0.2$ plus lensing. The $\ell=325.5$ bin does not pass null tests. }
\label{fig:pspec}
\end{figure}

In our implementation of MASTER we compute two different error bars for use with different tests. The ``MC errors'' (for Monte Carlo) are based on 400 simulations of the analysis pipeline except where noted. The other are termed ``LF'' for likelihood function.  For these, MC realizations are used to determine parameters for the likelihood functions of the binned bandpowers $\hat{C}_b$ as:
\begin{equation}
  \hat{C}_b \equiv  F_b^{-1} \sum_{b'} M^{-1}_{bb'} \tilde{C}_b \equiv {F_b}^{-1} \tilde{\tilde{C}}_b ,
\end{equation}
where $M_{bb'}$ includes the correction for the beam window function and the coupling matrix in Eq.~(\ref{eq:61}). The transfer function $F_b$ is obtained through the MC simulations.
These errors are used for analyses involving EE and BB spectra. The spectra are shown in Figure~\ref{fig:pspec} and given in Table~\ref{tab:pspec}.

\begin {table}[b!]
\centering
\caption{The ABS power spectra and cross spectra, ${\cal D}_\ell^{XX} \equiv C_\ell^{XX} \ell (\ell+1)/2\pi$. For the temperature in the TE and TB we use Planck's reobserved 143 GHz map and quote MC error bars. Similarly, for EB we give MC errors.  For EE and BB we quote maximum likelihood errors. The sample variance for EE ranges from 0.027 $\mu$K$^2$ in the first bin to 1.1 $\mu$K$^2$ in the last. Thus EE is dominated by measurement noise.  The bin-by-bin correlation in EE and BB spectra is only significant between nearest neighbors, where the correlation is $-7.5$\% on average.  All values are in $\mu$K$^2$.}
\label{tab:pspec}
\newcommand{\Tblh}{\rule[-2mm]{0mm}{7mm}}
\begin {tabular} {ccrrrrr}
  \hline\hline
\Tblh  $\ell$ center & $\ell$ range & \multicolumn{1}{c}{${\cal D}_\ell^{TE}$} & \multicolumn{1}{c}{${\cal D}_\ell^{TB}$} & \multicolumn{1}{c}{${\cal D}_\ell^{EE}$} & \multicolumn{1}{c}{${\cal D}_\ell^{BB}$} & \multicolumn{1}{c}{${\cal D}_\ell^{EB}$} \\\hline
\Tblh     55.5 &     41--70 & $ -3.8 \pm   2.5$ & $ -3.0 \pm   2.1$ & $ +0.33^{+0.10}_{-0.09}$ & $ +0.06^{+0.06}_{-0.05}$ & $+0.10 \pm  0.04$ \\
\Tblh     85.5 &  \ 71--100 & $-14.8 \pm   3.9$ & $ +6.1 \pm   3.0$ & $ +0.47^{+0.15}_{-0.13}$ & $ -0.03^{+0.08}_{-0.07}$ & $-0.15 \pm  0.08$ \\
\Tblh    115.5 &   101--130 & $  -28 \pm     6$ & $   -7 \pm     5$ & $ +0.97^{+0.24}_{-0.21}$ & $ +0.07^{+0.13}_{-0.12}$ & $+0.01 \pm  0.12$ \\
\Tblh    145.5 &   131--160 & $  -46 \pm     7$ & $    0 \pm     6$ & $ +0.59^{+0.25}_{-0.22}$ & $ +0.13^{+0.21}_{-0.19}$ & $-0.08 \pm  0.17$ \\
\Tblh    175.5 &   161--190 & $  -25 \pm     9$ & $   +1 \pm     7$ & $ +0.25^{+0.30}_{-0.28}$ & $ +0.21^{+0.32}_{-0.28}$ & $+0.19 \pm  0.24$ \\
\hline
\Tblh    205.5 &   191--220 & $  -14 \pm    10$ & $   -4 \pm     9$ & $    +0.5^{+0.5}_{-0.4}$ & $    -0.2^{+0.4}_{-0.4}$ & $ +0.1 \pm   0.3$ \\
\Tblh    235.5 &   221--250 & $  +42 \pm    12$ & $  -16 \pm    11$ & $    +1.2^{+0.7}_{-0.7}$ & $    -0.5^{+0.6}_{-0.5}$ & $ -1.0 \pm   0.5$ \\
\Tblh    265.5 &   251--280 & $  +69 \pm    13$ & $  +13 \pm    11$ & $    +2.3^{+1.1}_{-1.1}$ & $    -0.3^{+0.9}_{-0.8}$ & $ +1.0 \pm   0.8$ \\
\Tblh    295.5 &   281--310 & $  +76 \pm    15$ & $  -12 \pm    13$ & $    +5.1^{+1.7}_{-1.6}$ & $    +0.1^{+1.5}_{-1.4}$ & $ -0.5 \pm   1.1$ \\
\Tblh    325.5 &   311--340 & $ +118 \pm    17$ & $   -1 \pm    13$ & $   +17.2^{+3.4}_{-3.2}$ & $    +5.3^{+2.4}_{-2.3}$ & $ +1.9 \pm   2.0$ \\
\hline
\Tblh    355.5 &   341--370 & $  +40 \pm    16$ & $   -6 \pm    15$ & $   +10.3^{+4.2}_{-3.9}$ & $    -0.9^{+3.4}_{-3.2}$ & $ +4.3 \pm   2.7$ \\
\Tblh    385.5 &   371--400 & $  +24 \pm    17$ & $  -11 \pm    16$ & $         +14^{+6}_{-6}$ & $          -3^{+5}_{-5}$ & $   -1 \pm     4$ \\
\Tblh    415.5 &   401--430 & $  -51 \pm    21$ & $  -37 \pm    18$ & $        +21^{+10}_{-9}$ & $          -7^{+8}_{-8}$ & $   +7 \pm     7$ \\
\hline
\end {tabular}
\end {table}

Simulations are an essential part of the analysis and used in every step. For example, to assess the effects of processing, cuts, and filtering,  signal-only simulations are run to estimate the binned transfer function $F_b$ and confirm the unbiased nature of the $\hat{C}_b$  estimator. We also run simulations that include the CMB signal and experimental noise (see \S\ref{sec:noise_model}) for computing the likelihoods used in \S\ref{sec:r_limit} and the null nests described below.

From the null tests, we found it necessary to filter out  $a_{lm}$ coefficients with $\ell \leq 70$, the first bin, and $ m \leq 4$ to eliminate spurious large-scale polarized power in null maps. This is a region where little sky signal survives the other timestream-level filters, and the additional effect of the filter in the $(\ell,m)$ space on signal is negligible.  We emphasize that the CMB spectra were blinded when the filter parameters were set.

\subsubsection{Analysis validation and null tests}
\label{sec:nulltests}

Null tests are an integral part of the analysis. We split the time ordered data into two subsets, $d_1$ and $d_2$, pass them through the entire pipeline, and then make maps for both the subsets. Since the CMB signal is the same in both, the differenced map should be a ``null map"
and the resulting power spectrum,  $\hat{C}_{b,null}^{data}$,  should be consistent with noise model.
We compare $\hat{C}_{b,null}^{data}$ with the expectation from full pipeline simulations for the same data split, $\hat{C}_{b,null}^{sims}$. This MC ensemble accounts for our noise model including non-white noise, correlation among bins, different effects of the filters on the two halves, and the correlation among different null splits for the statistical assessment of the null test suite.

\begin{figure}[t!]
\centering
\includegraphics[,clip=]{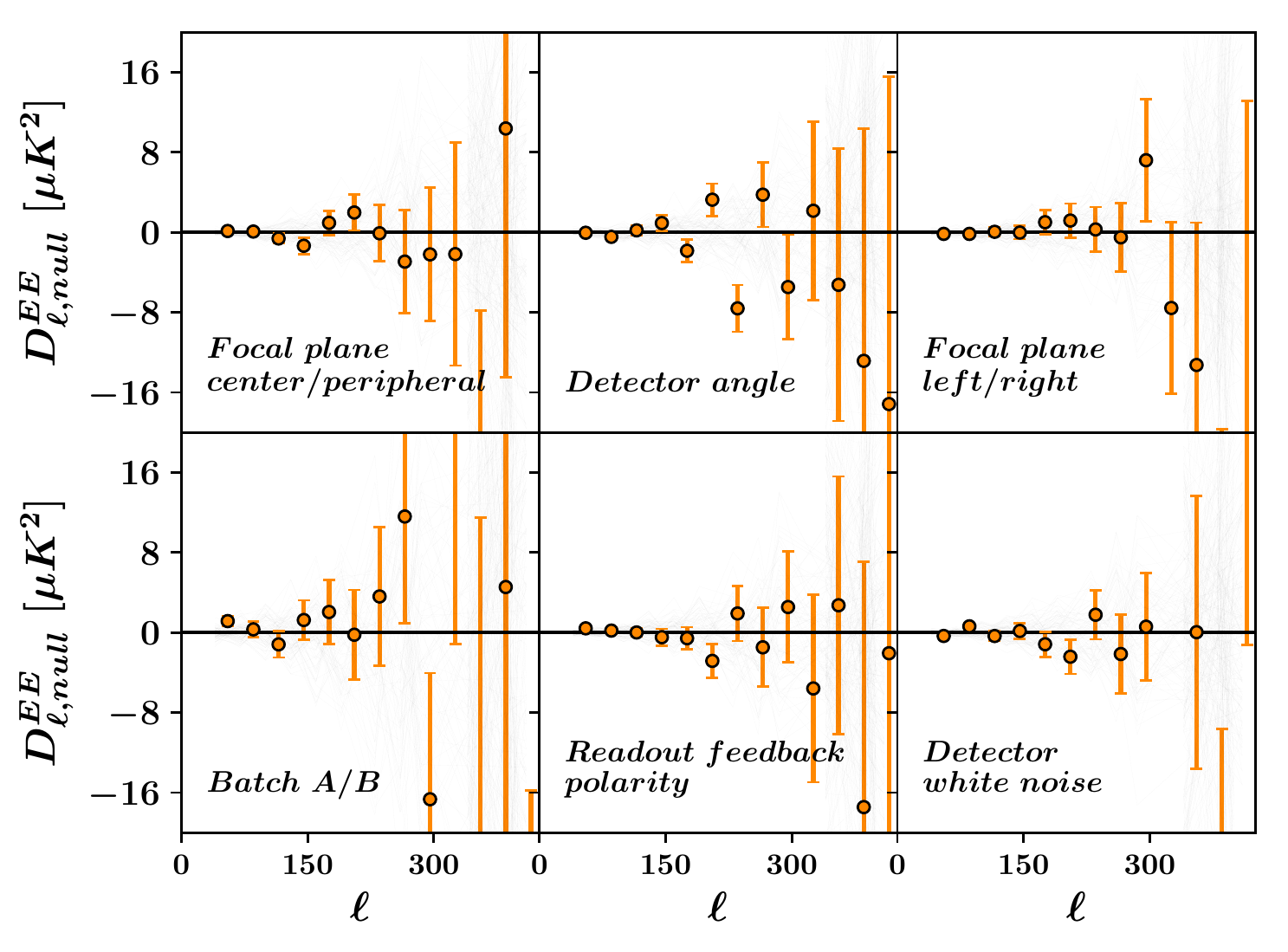}
\caption{Null tests for the EE power spectra with data splits based on detector performance. The orange circles represent the data points ${\cal \hat{D}}_{\ell, null}^{data}$. For comparison, the expectation spectra from the MC simulations ${\cal \hat{D}}_{\ell,null}^{sims}$ are shown as black semi-transparent lines. 
The $\ell$ = 325.5 data point for the white-noise null test is outside the y-axis range indicating its failure.}
\label{fig:ee_null_test_examples}
\end{figure}

We performed 21 null tests as shown in Table~\ref{tab:nullsuite_results} and analyzed the results using expectation
spectra obtained from 400 MC simulations. We calculated the null power $C_{b,null}^{data}$ from the differenced maps as
\begin{equation}
\hat{C}_{b,null}^{XY} = \frac{\tilde{\tilde{C}}_{b}(d_1)^{XY}}{F_{b}(d_1)}+\frac{\tilde{\tilde{C}}_{b}(d_2)^{XY}}{F_{b}(d_2)}-2 \frac{\tilde{\tilde{C}}_{b}(d_1,d_2)^{XY}}{F_{b}(d_1,d_2)},
\end{equation}
where XY is either EE/BB/EB and $F_b$ is the transfer function. We adopted a blind analysis strategy to avoid bias by not revealing the EE, BB, EB, and TB non-null spectrum until all the calibrations and data selection stages were finalized and the null tests were successful. Figure~\ref{fig:ee_null_test_examples} shows the EE null spectra for six null tests based on the performance of detectors.

\begin {table}[b!]
\centering
\caption{The ABS null suite contains 21 tests to probe the systematics in the data for $\ell<310$.  The $\chi^2_{null}$ calculated from nine $\ell$ bins and the corresponding PTE values are shown for each null test. The PTE for the detector angle split for EE is unusually low but we cannot identify a systematic that would produce this. For all EE and BB tests combined we find $\chi^2 = 393.8$ for $\rm{dof} = 378$ resulting in $\rm{PTE} = 0.28$. Additional details for each test are given in appendix~\ref{appendix_ABS_null_suite}. }
\label{tab:nullsuite_results}
\begin {tabular} {crrrrrrrrr}
  \hline\hline
  \multirow{2}{*}{Null suite type} && \multicolumn{2}{c}{EE (dof=9)} && \multicolumn{2}{c}{BB (dof=9)}  && \multicolumn{2}{c}{EB (dof=9)} \\
  && $\chi^{2}_{null}$ & PTE && $\chi^{2}_{null}$ & PTE && $\chi^{2}_{null}$ &PTE \\\hline
  	       \multicolumn{8}{l}{{1. Data quality}}\\
             Detector white noise &  &  13.3 &  0.14 &  &   9.7 &  0.39 &  &  15.5 &  0.11 \\
                     Glitch count &  &   6.3 &  0.72 &  &   8.9 &  0.47 &  &  12.9 &  0.17 \\
           Knee frequency &  &   9.3 &  0.41 &  &   3.6 &  0.92 &  &  10.5 &  0.34 \\
          Scan synchronous signal &  &   8.7 &  0.45 &  &  11.2 &  0.29 &  &   4.9 &  0.85 \\
         Statistical stationarity &  &  13.0 &  0.20 &  &   5.5 &  0.78 &  &  11.7 &  0.29 \\ \hline

  	       \multicolumn{8}{l}{{2. Instrument performance}}\\
                        Batch A/B &  &  10.3 &  0.30 &  &  11.2 &  0.25 &  &   3.8 &  0.93 \\
               Central/peripheral &  &   7.1 &  0.65 &  &   9.1 &  0.44 &  &   9.1 &  0.44 \\
                   Detector angle &  &  23.3 &  0.01 &  &  20.6 &  0.02 &  &  15.4 &  0.07 \\
           Focal plane left/right &  &   3.6 &  0.94 &  &  15.6 &  0.07 &  &   6.8 &  0.66 \\
                  HWP performance &  &  11.4 &  0.24 &  &   8.9 &  0.43 &  &   7.0 &  0.60 \\
        Readout feedback polarity &  &  10.3 &  0.35 &  &   9.4 &  0.41 &  &   9.9 &  0.34 \\ \hline

  	       \multicolumn{8}{l}{{ 3. Observing conditions}}\\
              Ambient temperature &  &  12.6 &  0.14 &  &  14.0 &  0.12 &  &  14.1 &  0.14 \\
          Azimuth east/west scans &  &  13.6 &  0.15 &  &   2.8 &  0.99 &  &   4.1 &  0.95 \\
                         Humidity &  &   7.1 &  0.61 &  &   8.1 &  0.52 &  &  17.6 &  0.06 \\
                              PWV &  &   9.1 &  0.41 &  &  13.3 &  0.18 &  &  13.2 &  0.15 \\
                       Wind speed &  &   6.9 &  0.65 &  &   7.1 &  0.61 &  &  16.9 &  0.04 \\ \hline

  	       \multicolumn{8}{l}{{ 4. Temporal variations}}\\
                    Chronological &  &   8.7 &  0.45 &  &   4.4 &  0.88 &  &   3.7 &  0.92 \\
         Moon above/below horizon &  &   9.5 &  0.40 &  &   9.4 &  0.39 &  &  11.3 &  0.26 \\
                    Moon distance &  &   3.5 &  0.94 &  &   7.7 &  0.58 &  &   2.5 &  0.99 \\
          Sun above/below horizon &  &  10.0 &  0.32 &  &   8.5 &  0.44 &  &   9.3 &  0.41 \\
                     Sun distance &  &   4.5 &  0.86 &  &   2.8 &  0.97 &  &  11.0 &  0.25 \\  \hline  \hline
                   Total &  & {202.0} & { 0.28} &  & {191.8} & { 0.44} &  & {211.2} & { 0.13} \\ \hline

\end {tabular}
\end {table}

\begin{figure}[!t]
\centering
\includegraphics[width=1.\textwidth, height=0.9\textwidth,clip=]{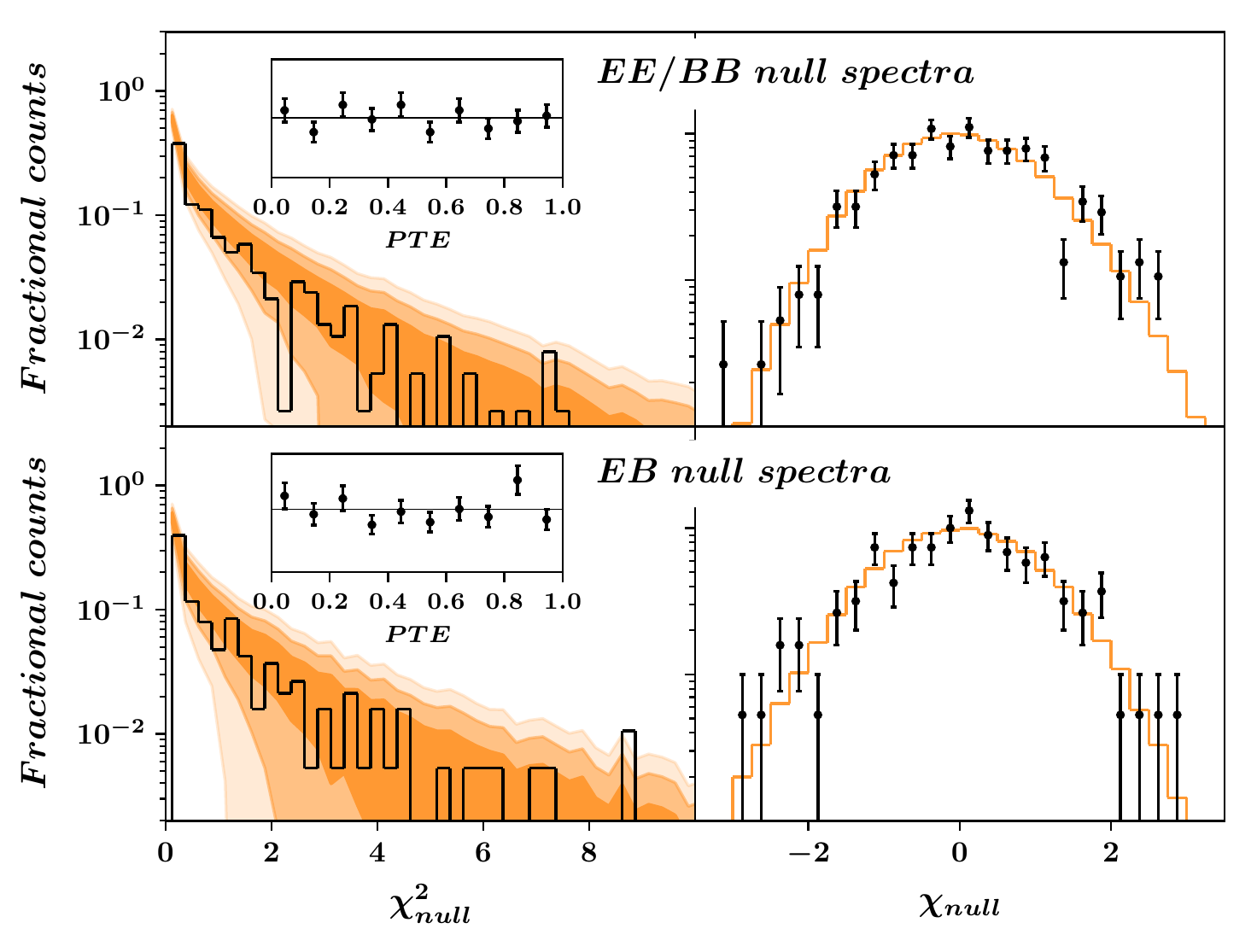}
\caption{The black histogram in the top left panel shows the distribution of $\chi^{2}_{null}$ values of the data for the nine power spectral bins for the EE/BB power spectra for the 21 null tests in the ABS null suite. The data points lie well between the ($1\sigma,\ 2\sigma$, and $3\sigma$) errors obtained from MC simulations which are shown as different shades of orange around the black histogram. The inset plot shows the uniform distribution of the PTE values indicating the success of the null tests. The overall PTE value of 0.28 is within the expected range. On the top right is the distribution of $\chi_{null}$ values of the data (black) distributed well around the MC simulations shown as the orange histogram. The bottom panels correspond to the $\chi^{2}_{null}, \chi_{null}$ distributions of the EB null test after opening the box in our blind analysis strategy. The distributions show no significant problems in the data.}
\label{fig:chisq_dist}
\end{figure}

We examine both $\chi_{null} =  \hat{C}_{b,null}/\sigma_b$  and $\chi^2_{null} $ for all the null tests in nine power-spectrum bins in the range $41<\ell \leq 310$, following \cite{quiet:2011, srini_thesis}.  The weighting factor $\sigma_b$ is the standard deviation of the 400 MC simulations, $C_{b,null}^{sims}$, in the respective bin. We assessed the $\chi^2_{null}$ statistics by calculating the probabilities to exceed (PTE), defined as the percentage of MC simulations that have $\chi^2_{null,sims}>\chi^2_{null,data}$. The PTE values are expected to follow a uniform distribution between zero and unity if the differenced maps are consistent with the null result. 
The distribution of $\chi^2_{null}$ are shown in left panels of Figure~\ref{fig:chisq_dist}. The corresponding $1\sigma$, $2\sigma$, $3\sigma$ errors, shown as different shades of orange, are calculated using MC simulations. As is evident from the figure, the the null tests are consistent with noise. 
The sum of the $\chi^2_{null}$ for the total of $2 ({\rm EE/BB}) \times 9 (\ell-{\rm bins})\times 21 ({\rm null~tests})= 378$ ($9 \times 21 = 189$) bins is 394 (211) for EE+BB (EB) null spectra, corresponding to a PTE of 0.28 (0.13).
The $\chi_{null}$ distribution, which can capture the direction of the systematic bias in the data, is shown in the right panels of the Figure~\ref{fig:chisq_dist}. The distribution is well centered at zero with a mean value of $0.075\pm0.052$ for EE+BB and 
$0.028\pm0.076$ for EB null spectra. These results indicate that systematic bias is not significant in our data.

The earlier version of this test examined all 13 $\ell$-space bins. We found that the $\ell=325.5$ bin failed  the null test based on the detector white-noise levels.  The data entering it was inconsistent with our model.  The significance of the null power excess as a single bin is $4.1\sigma$, and the PTE to have such an excess out of the entire EE null test suite is less than 4\%.  We are not aware of the source of the failure.
This failure was found in the null test validation before ``opening the box.''  It is clearly an outlier even in non-null spectra as can be seen in EE and BB plots in the right panels of Figure~\ref{fig:pspec}.   
In the following, we present results for the first nine bins excluding all the bins at equal or higher $\ell$ than the failure, all thirteen bins, and just twelve bins excluding the $\ell$ bin of the failure. The selection of bins does not change any qualitative conclusions.

\subsubsection{Possible systematics effects}
\label{sec:systematics}

In estimating the systematic errors, we use two methodologies: propagating systematic effects through the full simulation pipeline and propagating possible uncertainties in the calibration constants by processing real data through the full pipeline while varying the constants and comparing the results, all while still blinded. With the latter method, variations in the calibration constants lead to slightly different statistical weighting of the data, and the difference before and after the variations includes a fraction of statistical error that is already accounted for in the error estimates. Thus, the systematic errors estimated in this manner are upper limits. In the high $\ell$ region, these residual statistical fluctuations grow and dominate the systematic error estimates. 

\begin{figure}[t!]
  \begin{center}
  \includegraphics[width=0.8\textwidth]{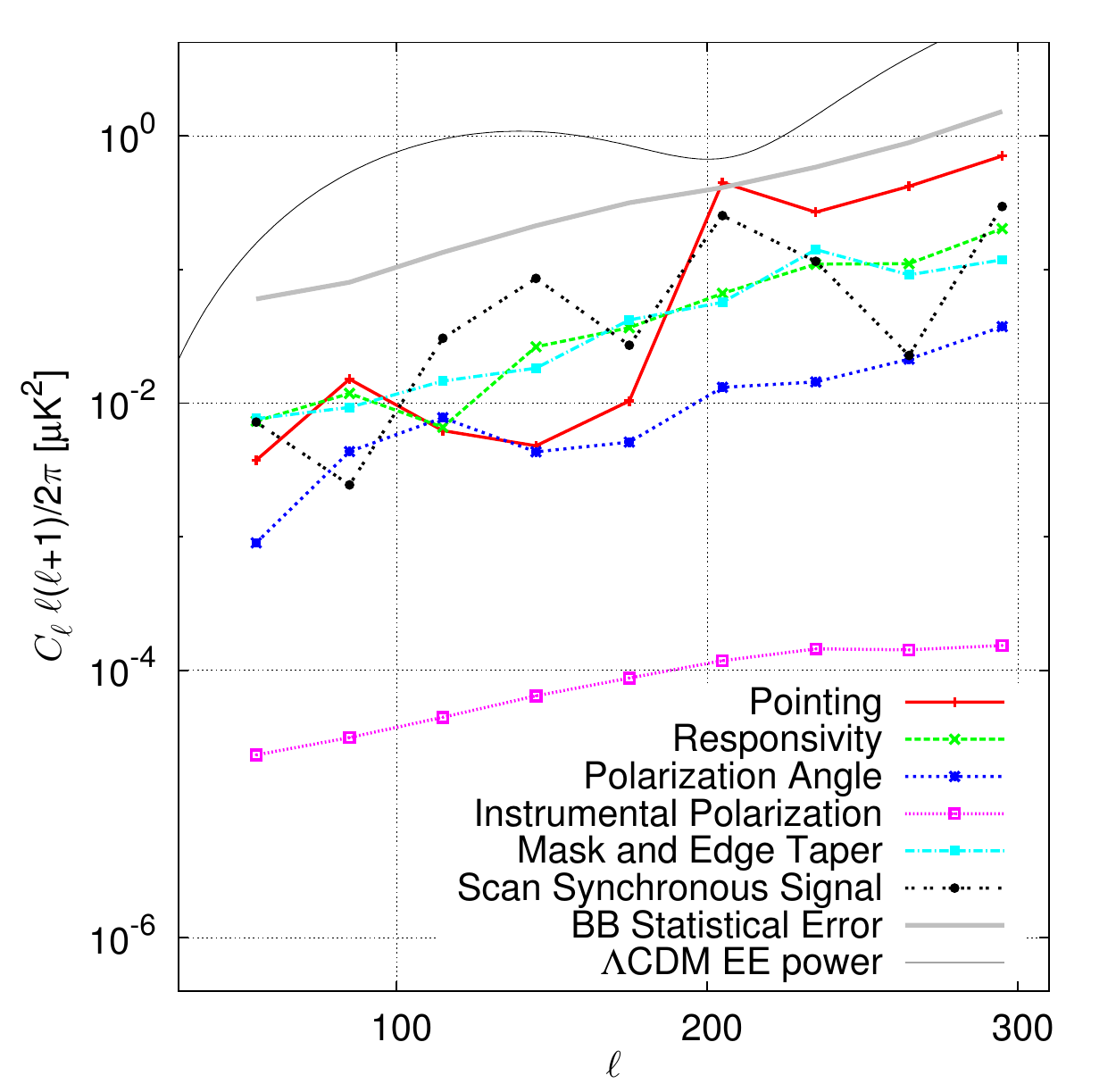}
  \caption{\label{fig:syst} 
    Systematic unceratinty estimates for the BB power spectrum.  Within each category, the errors are added in quadrature since the estimates do not have a preferred direction of bias.  Except for instrumental polarization and the polarization angle, the estimates are dominated by residual statistical fluctuations and are thus conservative upper limits.  The systematic uncertainties are well below the statistical uncertainty for $\ell<150$. 
  }
  \end{center}
\end{figure}

Multiple sources of systematic errors are grouped into categories. For each source, the bias has no preferential direction, so we add the systematic errors within each category in quadrature (with the exception of systematic errors from instrumental polarization). The power spectra of the systematic errors are shown in Figure~\ref{fig:syst} and are well below the statistical uncertainties for $\ell < 150$. We next describe the systematic error estimations for the sources in each category.

{\it Pointing uncertainty}.
The pointing uncertainty may lead to spurious B-mode power.  To estimate this effect, we evaluate two independent pointing models using different analytic forms.  We process data using the fiducial and alternative pointing models, and assign the difference as the systematic error.

{\it Responsivity}.
Possible uncertainties in the responsivity model lead to distortion in the map and may cause spurious signals.  Our responsivity model captures the declining trend of the responsivity of B-batch detectors over the observations as discussed in Sec.~\ref{sec:responsivity}.  To assess the possible systematic error due to this effect,
we use an alternative model where the responsivity is set to be constant across the entire data set for each detector. We process the data using the fiducial and alternative responsivity models, and assign the difference as the systematic error. We note that the treatment here is conservative since the actual uncertainty of the model is likely to be smaller than the difference between the fiducial and alternative models described here. As discussed, we have reconstructed the time trend of the responsivity not only using the $2f$ signal of the HWP, which is used in the fiducial model, but also using the relation between the PWV and the detector bias power.  The trend is consistent between the two methods, and the difference is smaller than the difference between the fiducial responsivity model and the constant-responsivity alternative model.

The relative responsivity among detectors is another aspect that comprises our responsivity model.  We randomly vary the relative responsivity by the amount of the calibration uncertainty, process the data using the
randomly-varied relative responsivity, compare the resultant power spectrum with the one using the fiducial responsivity, and assign the difference as the systematic error.

{\it Polarization Angles}.
The polarization angle model can be factored into three aspects: 1) the absolute detector angle, 2) the relative angle among the detectors, and 3) the time variation due to the combination of the continuously-rotating HWP and the variation in the detector time constant.  The absolute angle uncertainty is $\pm 1.1^\circ$. We estimate the possible spurious B-mode due to the possible variation of the absolute angle by this amount using signal-only simulations. The systematic error due to the relative angle among detectors is assigned by
randomly varying the relative angles by the calibration uncertainties, repeating the processing on signal-only simulations, and comparing with and without the variation. Similarly, the systematic error due to the uncertainty in modeling the time variation of the detector angles is assigned by varying the modeling parameters by their calibration uncertainties and processing signal-only simulations to estimate their impact on the final $BB$
spectrum.

{\it Instrumental polarization}.
Instrumental polarization, or so-called intensity-to-polarization leakage, creates spurious polarization
signals from intensity fluctuations.  The leakage can be decomposed into monopole and higher order terms such as the dipole 
and quadrupole~\cite{shimon/etal:2008}.  In our previous publication~\cite{essinger-hileman/etal:2015}, we described the details of this systematic error estimate and have shown that the continuously-rotating HWP mitigates this systematic. The ``monopole'' intensity-to-polarization leakage is 0.013\%, and the upper limit we set on the ``dipole'' and ``quadrupole'' intensity-to-polarization leakages are both 0.07\%.

{\it Point source masks and weight function}.
Point sources with an intensity flux larger than 1\,Jy are masked in our fiducial analysis (\S\ref{sec:tod2ps}).  We assess possible residuals to the point source contribution by comparing an analysis with the threshold of 400\,mJy.  We also estimate the integrated contribution of the unresolved sources to be small, with the possible bias corresponding to $r<0.01$ at $\ell<100$ \cite{dezotti/etal:2005,battye/etal:2011}.  In our fiducial analysis, the point source mask has a profile with a circle of radius of $0.25^\circ$ that is completely masked and then gradually tapered up to $1.5^\circ$ in radius. We confirm the mask profile has negligible impact by varying the mask and tapering radii by $0.05^\circ$ and $0.25^\circ$, respectively.

We apply an $\ell$-dependent weight function that combines the map noise level and expected signal power (LCDM for $C_\ell^{EE}$ and zero for $C_\ell^{BB}$) following the FKP ansatz~\cite{feldman/etal:1994,smith/etal:2007}.  Additionally, in order to remove the perimeter of Field A where the statistical contribution is negligible, we mask the lowest hit regions around the perimeter then smooth the sharp boundary with a Gaussian FWHM of $2^\circ$.  This results in $\sim 1$\% loss in the total effective sky area.  We compare the cases with and without the peripheral apodization and confirm the systematic impact of this mask is negligible.

{\it Variation in Scan Synchronous Signal}.
We assume that the scan synchronous signal (SSS) does not vary as a function of time during a CES ($\sim$1 hour); our pipeline only filters a SSS component that is constant over a CES for each TES and our MC simulation does not have a varying SSS as part of its noise model.  We cut the CES TES timestreams with a possible variation of the SSS within a CES as discussed in \S\ref{sec:SSS}.  However, comparison of the SSS among different CESes indicates residual non-zero variation on longer timescales such as hours and days.  Characterizing the timescale of the variation for each detector and interpolating it to a shorter timescale of $\sim1$\,hour, we identify $\sim15\%$ of detectors that may have a SSS variation within a CES that is potentially non-negligible.  In order to assess the effect of the possible variation of SSS, we compare the results with and without these 
$\sim15\%$ removed in the data selection.  The difference of the results show no significant systematic difference.  We note that this approach is conservative because there are features that can cause inter-CES variations but not intra-CES variations; our systematics estimate is based on the former, while the systematic errors would arise only from the latter.

\subsection{Comparison to Planck}
\label{sec:cfPlanck}
After the ABS spectra and null tests were complete, we cross correlated the ABS polarization maps with the Planck temperature  and polarization maps. The cross spectra are computed using two methods. In the first, we simply take the Planck value in each ABS pixel. In the second, we run the Planck maps through our observing strategy and mapmaking process. We then compute the auto and cross spectra shown in Figure~\ref{fig:pspec}. Although there are some differences between the two methods they are not large. In what follows we use the spectra from the second method.

\paragraph{\it Do ABS and Planck agree?} 
We first compute the calibration factor by finding the best fit between the spectra of the ABS and Planck maps. If $\alpha$ is the factor that divides the ABS map for calibration to Planck, then we can estimate $\alpha$ with $\alpha = {\cal D}^{AA}/{\cal D}^{AP}$, where $A$ is the ABS map, $P$ is the 143 GHz reobserved Planck map, and ${\cal D}^{XY}$ is the EE power spectrum estimated with maps $X$ and $Y$. We use 80 MC simulations each for the two spectra (common signal for $A$ and $P$ but each with its representative noise simulation), then compute the statistical uncertainty on $\alpha$ for each bin from the rms of the simulations. Taking the weighted mean over all bins (all but the 10th bin, which we know is unreliable from null tests), we find $\alpha = 0.89 \pm 0.10$ ($\alpha = 0.84 \pm 0.10$). Because the uncertainty on the fit is larger than that from the planet calibration, we take the planet calibration as definitive. In the following, the ABS and Planck calibrations are fixed at their nominal values.

We proceed to evaluate the power spectrum of the null map, $A - P$, and assess the significance with MC simulations, following the procedure as in the ABS internal null tests described in Section~\ref{sec:nulltests}. The power spectrum of the null map is given by,
\begin{equation} 
{\cal D}_b^{A-P} =  {\cal D}_b^{AA} +  {\cal D}_b^{PP} - 2  {\cal D}_b^{AP},
\label{eq:null_spec}
\end{equation}
where, in practice, we evaluate each power spectrum on the right hand side of equation~\ref{eq:null_spec} using cross spectra between the corresponding data splits to avoid noise bias. We estimate the uncertainty $\Delta {\cal D}_b^{A-P}$ using 100 noise only MC simulations. Using noise only simulations assumes the signals in the two maps cancel, and hence the fluctuation of the null power spectrum should be consistent with the statistical uncertainties in the two maps. For ABS, we generate the simulations from the time-ordered-data and thus they include all the data processing. For Planck we use their simulations (FFP8) and reobserve them following the ABS coverage. For the thirteen bins in EE, we find $\chi^2/\nu=11.4/13$ with a ${\rm PTE=0.5}$, where $\chi^2 =  \sum_b({\cal D}_b^{A-P}/\Delta {\cal D}_b^{A-P})^2$. The first bin is $1.8\sigma$ above zero but statistically this is not unexpected as shown by the PTE. The distribution is well fit by a $\chi^2$ distribution with thirteen degrees of freedom.

We conclude that the ABS and Planck maps are consistent to the limits of noise. Or, in the power spectra the systematic errors are subdominant to any statistical errors. 

\subsection{Foreground emission}
\label{sec:foregrounds}
There are two well-established sources of polarized foreground emission. They are synchrotron emission and thermal dust emission. At the current level of precision, they may be modeled as
\begin{equation}
\begin{split}
M(\nu) &= S(\nu) + D(\nu) \\
&= \alpha_s(\nu /\nu_K)^{\beta_s}\mathbb{S} + \alpha_d {\cal E}_d(\nu,\beta_d)\mathbb{D},\\
\end{split}
\end{equation}
where $M(\nu)$ is a Stokes Q or U map in antenna temperature at frequency $\nu$. The quantities $\mathbb{S}$ and $\mathbb{D}$ are normalized synchrotron and dust spatial templates at frequencies 
$\nu_K=22.3$\,GHz (WMAP) and 353\,GHz (Planck), with amplitudes $\alpha_s$ and $\alpha_d$. The dust emission is described  as \cite{planck_dust:2015}
\begin{equation}
{\cal E}_d(\nu,\beta_d)\equiv\biggl(\frac{\nu}{\nu_{\rm ref}} \biggr)^{\beta_d-2}
\frac{B_\nu(T_{dust})}{B_{\nu_{\rm ref}}(T_{dust})},
\end{equation}
where $B_\nu(T)$ is the Planck function and $\nu_{ref}=353$~GHz. For polarized dust emission we take $T_{dust}=19.6~$K and $\beta_d=1.59$ \cite{planck_dust:2015}. The polarized synchrotron and dust maps are correlated \cite{page/etal:2007,planck_dust:2015, choi/page:2015} but the effect is negligible for a single frequency at 145~GHz. Planck also found that the polarized synchrotron and dust scale as ${\cal D}_\ell\propto \ell^{-0.44}$ and $\propto \ell^{-0.42}$ respectively. We model the $\ell$ dependence with a pivot at $\ell=80$ \cite{planck_dust:2016, planck_fg:2016} for a straightforward comparison.

\begin{figure}[htbp]
  \begin{center}
  \includegraphics[width=\textwidth]{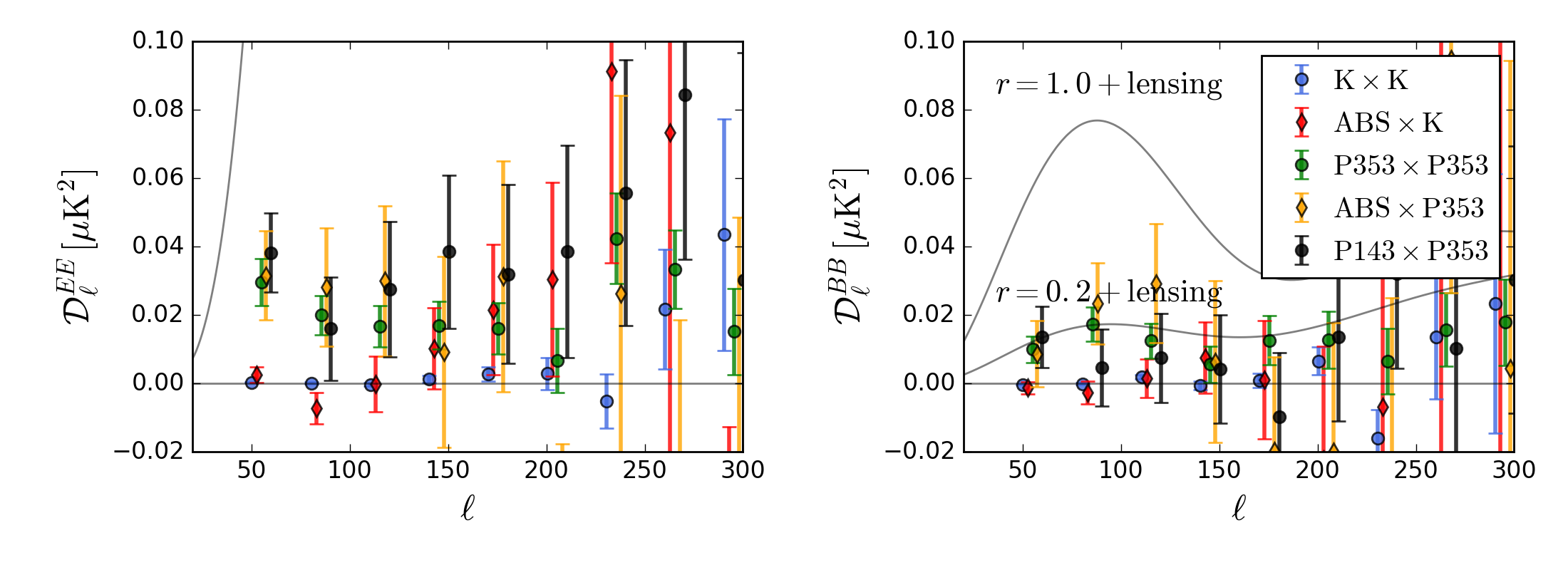}
  \caption{In each plot we show the power spectra of reobserved WMAP K-band and Planck 353~GHz maps. The spectra are scaled to 145~GHz, the ABS observing frequency. We also show in each plot the cross spectra of the maps with ABS, again scaled for 145~GHz. On the EE spectrum ({\it left}) we show the LCDM model and on the BB spectrum ({\it right}) we show $r=1$ and $r=0.2$ models.  It is clear by eye that the dust dominates and that ABS cross Planck is consistent with the extrapolated Planck spectrum. }
  \label{fig:fg_fig1}
  \end{center}
\end{figure}

\begin{figure}[htbp]
  \begin{center}
  \includegraphics[width=0.49\textwidth]{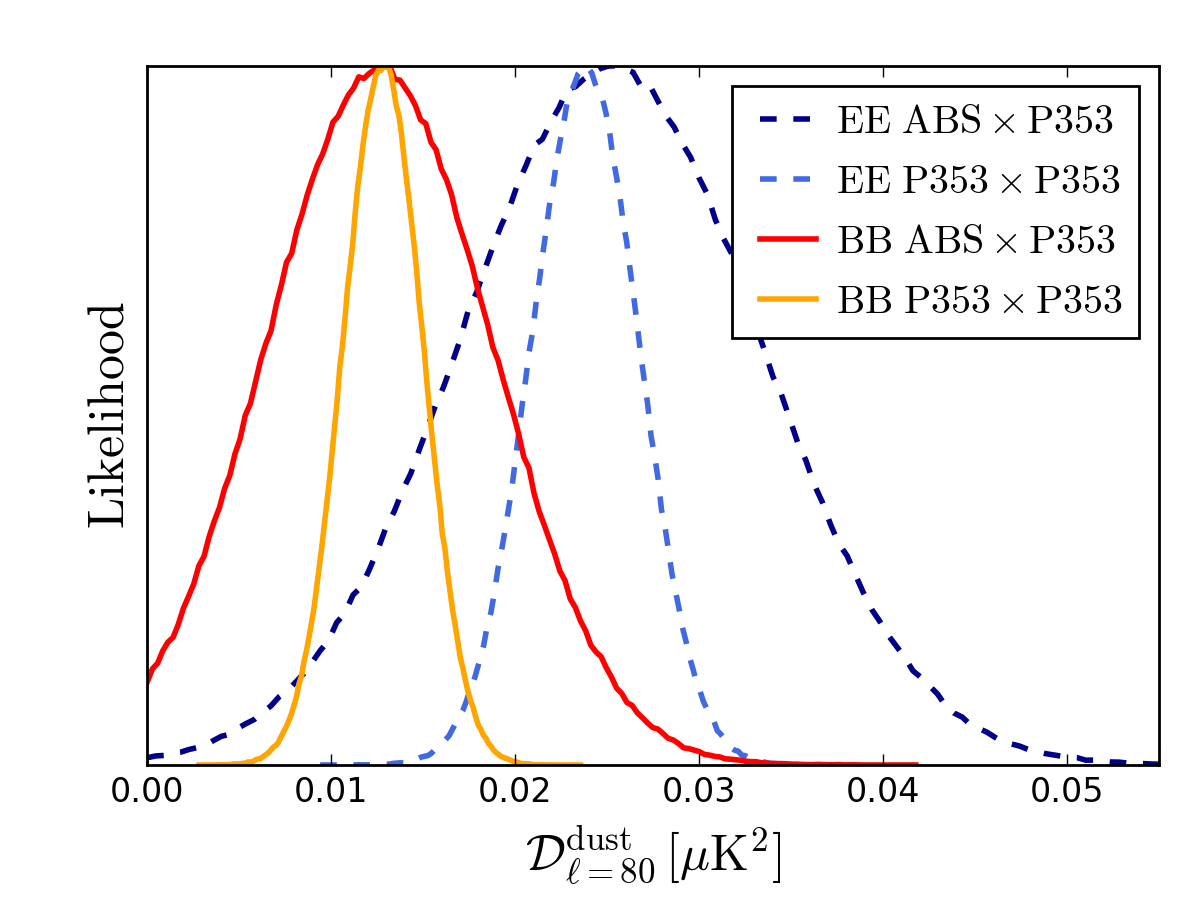}
   \includegraphics[width=0.49\textwidth]{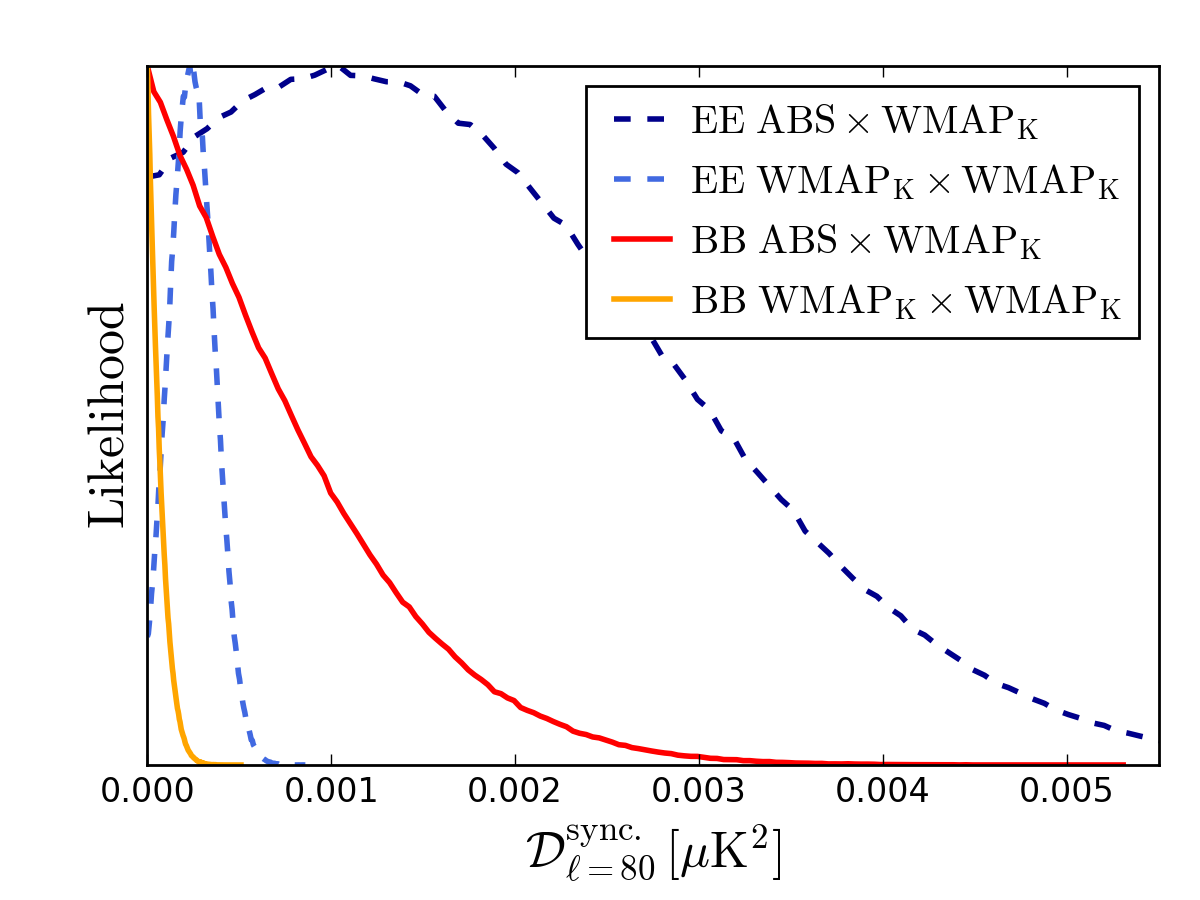}
  \caption{Each plot shows the likelihood of a model fit to the reobserved Planck 353 ({\it left}) and WMAP K band  ({\it right}) power spectra as well as the same model fit to the cross spectra with ABS. Both Planck and WMAP are scaled to 145~GHz. The models are ${\cal D}_{\ell=80}^{dust}(\ell/80)^{-0.42}$ and ${\cal D}_{\ell=80}^{sync.}(\ell/80)^{-0.44}$. The ABS calibration uncertainty is not included here. }
  \label{fig:fg_fig2}
  \end{center}
\end{figure}

As a first assessment of the level of foreground emission, we simply extrapolate the best fits to the power spectra of the reobserved WMAP K-band and Planck 353 GHz maps to 145~GHz. These are shown in Figure~\ref{fig:fg_fig1}. For synchrotron, we extrapolate with $\beta_s=-2.7$. There is no detected power at K band. We obtain limits at $\ell=80$ of 0.0005 $\mu$K$^2$(95\%cl) and 0.0002 $\mu$K$^2$ (95\%cl) for EE and BB respectively. The synchrotron spectral index appropriate for regions of low intensity at 145 GHz is not well constrained \cite{choi/page:2015}, with values of $-3.3<\beta_s<-2.5$ plausible.  For dust there is significant power detected in the 353 GHz spectrum. When extrapolated with $\beta_d=1.59\pm0.095$ to 145~GHz we find  $0.024\pm 0.003$ $\mu$K$^2$ and $0.013\pm 0.002$  $\mu$K$^2$ for EE and BB respectively. The uncertainty in $\beta_d$ is from
$\sigma_{\beta_{d}}=0.17(0.0076/f_{\rm sky})=0.095$  where  $f_{sky}=1000~$deg$^{2}$ \cite{planck_dust:2016}.  
For both extrapolations the contribution from the CMB is negligible. The results are shown in Figure~\ref{fig:fg_fig2}.

We also cross correlate ABS with the reobserved WMAP K-band and Planck 353~GHz maps. For synchrotron, a fit to the cross spectra yields  0.0042 $\mu$K$^2$  (95\%cl) and 0.002 $\mu$K$^2$  (95\%cl) for EE and BB respectively. For dust, we find $0.026\pm 0.008$ $\mu$K$^2$ and $0.013\pm 0.006$ $\mu$K$^2$  for EE and BB respectively. The cross spectrum is consistent with the extrapolation showing that ABS is sensitive to a small signal buried in the noise. It is noteworthy that even in this low-dust region, the BB/EE ratio is $\approx 0.5$ as seen in  \cite{page/etal:2007, gold/etal:2011, planck_dust:2016} in other regions. The level of dust emission is similar to that found in the BICEP2 region by Planck, $1.32^{+0.28}_{-0.24}\times10^{-2}\,\mu$K$^2$  where the uncertainty is from the extrapolation from 353 GHz to 150 GHz\cite{planck_dust:2016}. 

To compare to LCDM below we correct for the foreground in EE by simply subtracting the best fit model to Planck. To obtain an upper limit on $r$ we do not correct for foreground emission and simply quote an upper limit on the combination of foregrounds and any signal. The improvement from the subtraction is negligible.

\subsection{Agreement with LCDM}
\label{sec:lcdm} 
\begin{figure}[!ht]
\centering
\includegraphics[width=0.45\textwidth]{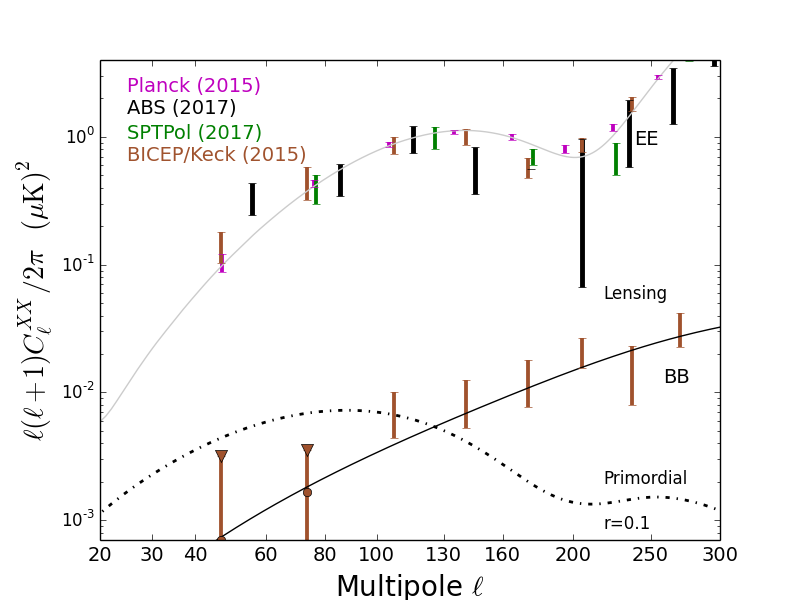}
  \includegraphics[width=0.45\textwidth]{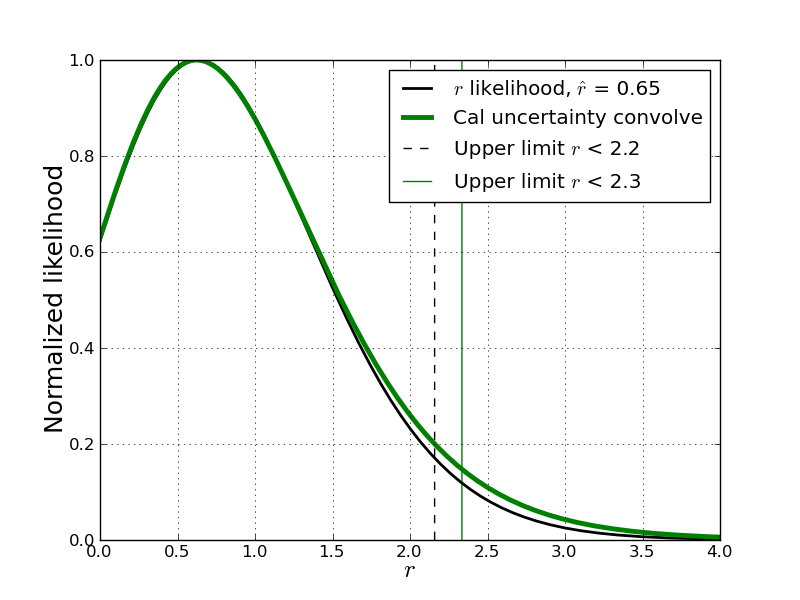}
\caption{{\it Left:} Recent two-point function measurements for $\ell<300$ from ABS,
Planck \cite{planck_overview:2015}, 
BICEP/Keck \cite{B2K-2016}, and SPT\cite{henning/etal:2017}. {\it Right:} The likelihood for $r$ from ABS. We find   $r<2.3$ (95\% cl)  including foregrounds and calibration uncertainty.}
\label{fig:ee_bb_spec}
\end{figure}

We assess the agreement through a simple $\chi^2$ fit of the power spectrum to the Planck model for EE. In the covariance matrix we include the calibration uncertainty, 6\%, and the beam uncertainty as in \S\ref{sec:beams}. We use the likelihood function (LF) error bars described in the next section. We also use LCDM simulations to find the effective $\ell$ for each band. These values differ by only 1-3 from the central values from 
Table~\ref{tab:pspec}. Excluding the $\ell=325.5$ bin, we find $\chi^2=19.3$ with a PTE$=0.08$; and considering just the first nine bins we find $\chi^2=15.1$ with a PTE$=0.09$. We conclude the data are consistent with LCDM. The EE spectrum is shown in Figure~\ref{fig:ee_bb_spec} along with other recent measurements.

For EB and TB we compare to a null signal with MC error bars for $\nu=9, 12$ (excluding $\ell=325$), and 13 bins. For EB we find
$\chi^2/\nu = 15.1/9, 18.7/12,\,{\rm and}\, 19.7/13$ with PTEs of 0.089, 0.095, and 0.11 respectively.  For TB we find
$\chi^2/\nu = 12.9/9, 17.6/12,\,{\rm and}\, 17.6/13$ with PTEs of 0.17, 0.13, and 0.17 respectively. We conclude that these too are consistent with LCDM.

\subsection{Limit on r}
\label{sec:r_limit}

To compute the upper limit on $r$ and the likelihood function error bars we follow the formalism adopted for QUIET \cite{quiet:2011}. 
We construct a likelihood function for computing the binned bandpowers using the MASTER pipeline of MC realizations. We model the probability distribution function of these variables with a two-parameter family of functions parameterized as a scaled $\chi^2$ function with degrees of freedom $\nu$ and standard deviation $\sigma$ centered on zero:
\begin{equation}
\label{eq:scaled_chi2}
P^M_{\chi^2}(x | \nu, \sigma) = \frac{\sqrt{2 \nu}}{\sigma} P_{\chi^2} \left( \nu \left[ \sqrt{\frac{2}{\nu}} \; x/ \sigma  + 1 \right] | \; \nu \right)
\end{equation}

This equation is used to describe the distribution of measured bandpowers (i.e. bandpowers recovered from MC realizations), $\hat{C}_b$, conditional on a theory bandpower $C_b$, as follows:
\begin{equation}
\label{bandpower_likelihood}
P(\hat{C}_b | C_b) = P^M_{\chi^2} \left( \frac{\hat{C}_b + N_b}{C_b + N_b} - 1 | \nu, \sigma \right) / (C_b + N_b)
\end{equation}
where here $N_b$ represents noise bias in the measurement, which is estimated from comparing fiducial MC realizations to the fiducial spectra used to generate them. During our fits of the PDF arising from MC simulations, we determined that reducing the PDF of Equation~\ref{eq:scaled_chi2} to the scaled $\chi^2$ distribution by setting $\sigma = \sqrt{ 2/ \nu }$ was sufficient to describe the distribution over MCs. We report bandpower errors based on the single-parameter likelihood, depending only on $\nu$, in Table~\ref{tab:pspec} and show these errors on the 
spectra in Figure~\ref{fig:pspec}.

Note that in this formulation, the uncertainty depends on the measured power as opposed to the expected power in the LCDM model. The BB spectrum is shown in Figure~\ref{fig:pspec}. There is clearly no measured signal. For all 13 bins, $\chi^2/\nu = 11.5/13$ with a PTE = 0.57. From the spectrum it appears that the $\ell=325.5$ is an outlier. This is not unexpected as this bin did not pass our null tests as discussed above. When excluded we 
find  $\chi^2/\nu = 5.2/12$ with a corresponding PTE of 0.96. We confirm this PTE with 400 MC simulations.

For $r$, the equation is similar:
\begin{equation}
\label{r_likelihood}
P(\hat{r} | r) = P^M_{\chi^2} \left( \frac{\hat{r} + n}{r + n} - 1 | \nu, \sigma \right) / (r + n)
\end{equation}
with $\hat{r}$ representing a measured $r$ value based on fitting to MC bandpowers, $r$ representing the theory value of the MC realizations, and $n$ representing the same effect in this formulation as $N_b$. 

We used two separate $r$ fitting pipelines for MC realizations: a minimum-$\chi^2$ using bandpower errors estimated from the MC ensemble and a maximum-likelihood method involving knowledge of the bandpower parameters ($\nu, \sigma$). For both methods we use a template that accounts for the bandpower window functions, beam effects, and other important instrumental factors. Both pipelines were checked with simulations and found to be free of bias when recovering $r$. All fitting is done with a pivot point of $k = 0.05$ Mpc$^{-1}$.

Figure~\ref{fig:ee_bb_spec} shows the result from the MC ensemble. Evaluated over positive values of $r$ we find $r<2.2$  (95\% cl) which rises to 
 $r<2.3$ (95\% cl) after accounting for calibration uncertainty following the method in \cite{ganga/etal:1997}. This limit includes a small contribution from dust emission. The upper limits for these three bins combined is ${\cal D}_l^{BB}<0.16~\mu$K$^2$ (95\% cl).  With the maximum likelihood pipeline we  perform a joint fit over the three parameters ($\nu, \sigma, n$) to determine errors given the value of $r$. We record $r = 0.65\pm 0.7 $, with the error bars representing 1$\sigma$ intervals.

For our nominal analysis, we use just the first three $\ell$ bins. This was decided on before the data were unblinded and was based on minimizing the potential of lower weight points not near the BB peak, biasing the result.  If we include the first four bins and repeat the analysis, the limit rises to  $r<2.4$ (95\% cl) including calibration error.

\section{Discussion and conclusions}
\label{sec:conc}

We have presented a new measurement of $r<2.3$ (95\% cl) based on direct measurements of the B-mode spectrum.  Over the past decade other direct B-mode limits have come from QUIET, with $r<2.8$ (95\% cl) \cite{quiet:2012}, and BICEP/Keck when combined with Planck, with $r<0.07$ (95\% cl)\cite{B2K-2016}.  The ABS instrument introduced the use of a cryogenic reflecting telescope for B-mode searches, was one of the first to field 
feedhorn-coupled TES bolometers, and perhaps most significantly, introduced the use of a rapid polarization modulator as the most skyward optical element. The spinning HWP led to a number of new analysis techniques. In addition to the stability it afforded, it provided a new way to continuously monitor the instrument responsivity, data quality and measure the time constants. 

Through cross correlation with Planck, ABS marginally detects the dust polarization in BB in a fairly clean part of the sky at levels comparable to those measured by B2K. We also presented observations of Tau-A taken with a rotating HWP. For mid-latitude observations, Tau-A holds promise as a calibration standard for the polarization angle. 

ABS is the first experiment to attempt observations of the large angular scale CMB polarization with bolometers at 145 GHz in Chile. The closest previous measurement in frequency was QUIET at 95 GHz \cite{quiet:2012}. QUIET used a correlation receiver. At 95 GHz the atmospheric fluctuations are 1/3 \cite{paine_scott_2017_438726} what they are at 145 GHz in Chile in the effective CMB temperature. Even so, the first $\ell$ bin for QUIET spanned from $25<\ell<75$ as compared to  $41<\ell<75$ with $\ell_{eff}=55$ for ABS.  Given the stability of the demodulated ABS data, one can in principle probe even larger scales. In the end, the large angular scale coverage was limited by the $7.1^{\circ}$ scan on the sky. The scale of the lowest mode the scan couples to is $\ell\approx 180^\circ/3.5^\circ\sim 50$ which is close to what we achieved. Future work will target larger scale scans.

ABS is a path-finder for future observations. Even though it is roughly an order of magnitude less sensitive than current CMB polarization experiments, it demonstrated control of systematic errors, particularly temperature to polarization leakage, that will be crucial in reaching $r<0.001$. Specifically, by direct measurement the monopole leakage term in ABS corresponds to a contamination of $r<0.001$ (95\% cl). The dipole and quadrupole leakage terms are not detected and correspond to an upper limit of $r<0.01$ (95\%cl)\cite{essinger-hileman/etal:2015}.
Given access to the large low-foreground area of the southern sky, the rotation of the sky to modulate the polarization relative to the horizon, the low atmospheric loading, and the year-round access with increasing infrastructure in place,  Chile is an excellent site for the search for B-modes.

\section{Acknowledgements}
\label{sec:ack}
The ABS experiment began in 2006.  This work was supported by the U.S. National Science Foundation through awards PHY-0355328, PHY-0855887, and PHY-1214379; and by the U.S. National Aeronautics and Space Administration through award NNX08AE03G.  Funding was also provided by Princeton University and a Canada Foundation for Innovation (CFI) award to UBC. ABS operated in the Parque Astron\'omico Atacama in northern Chile under the auspices of the Comisi\'on Nacional de Investigaci\'on Cient\'ifica y Tecnol\'ogica de Chile (CONICYT). Computations were performed on the GPC supercomputer at the SciNet HPC Consortium. SciNet is funded by the CFI under the auspices of Compute Canada, the Government of Ontario, the Ontario Research Fund Ð Research Excellence; and the University of Toronto. Work at LBNL is supported in part by the U.S. Department of Energy, Office of Science,  Office of High Energy Physics, under contract
No. DE-AC02-05CH11231.
TEH was supported by a National Defense Science and Engineering Graduate Fellowship, as well as a National Science Foundation Astronomy and Astrophysics Postdoctoral Fellowship. SMS and KC gratefully acknowledge support from NSTRF Fellowships.
SR acknowledges the CONICYT PhD studentship and support from Australian Research Council's Discovery Projects scheme (DP150103208). 
LEC acknowledges partial support from CONICYT Anillo project ACT-1122 and the Center of Excellence in Astrophysics and Associated Technologies (PFB06). AK acknowledges the support as a Dicke Fellow at Princeton University, and by JSPS KAKENHI Grant Number JP16K21744 and JSPS Leading Initiative for Excellent Young Researchers (LEADER). 
We thank the Mishrahi Fund and the Wilkinson Fund for generous support of the project.

Over the past decade, many have contributed to the building of, observing with, and analysis of ABS. They include Dick Bond at CITA, Pablo Gonzalez at UChile, Mark Halpern at UBC, Toby Marriage at JHU, and Ed Wollack at NASA/GSFC. They also include Princeton graduate students: Farzan Beroz, Emily Grace,  Silviu Pufu, and Sasha Zhiboedov; Princeton undergraduates: George Che, Dongwoo Chung, Ovidiu Cotlet, Rutuparna Das, Greg Dooley, Sean Frazier, Jonathon Goh, Kamna Gupta, Michael Jimenez, Alicia Kollar, Arjun Landes, Alex Leaf, Jennifer Lin, Cary Malkiewich, Peter Petrov, Jason Pollack, Dragos Potirniche, Cheryl Quah, Nicole Quah, Malik Rohan, Bogdan Stoica, Kai Sheng Tai, Peter Toshev, Evan Warner, Christine Wen, Xiaoweng Xu, Michael Zhang, and Tony Zhu; UChile undergraduate Basti\'an Pradenas; and visitors: Venus Appel, Charlotte Blais, Tyler Evans, and Alex Kinsey. We thank them for their efforts and insights.

The Princeton machine shop (Bill Dix, Glenn Atkinson, Fred Norton, and Laszlo Varga) was invaluable for building ABS as were Bert Harrop and Stan Chidzik in the elementary particle physics electronics group. We thank our many colleagues from ALMA, APEX, and \textsc{Polarbear} who helped us at critical junctures. Colleagues at RadioSky provide logistical support and kept operations in Chile running smoothly. 

ABS was located next to the Atacama Cosmology Telescope (ACT). ACT team members, especially Mark Devlin, Ben Schmitt, and Marius Lungu from the University of Pennsylvania and Bob Thornton at West Chester University, graciously helped us on numerous occasions. Masao Uehara, Rebecca Jackson, and Felipe Rojas worked tirelessly to keep ABS operating in Chile.

\appendix
\section{ABS null suite}
\label{appendix_ABS_null_suite}

Here we describe the ABS null test suite (Table \ref{tab:nullsuite_results}) consisting of 21 data splits designed to probe the systematics due to data quality, performance of the instrument, observing conditions, and temporal variations.
While each of the 21 null tests is designed to probe the specific systematic effect described below, there exist correlations between the different null tests thus they are not completely independent.  For example, the null tests based on the {\it position of Sun} is correlated to the {\it Ambient temperature} and {\it Chronological} null tests; and the null test based on PWV level is correlated with the {\it Knee frequency} and {\it HWP performance} null tests.  More details about the ABS null test suite can be found in \cite{srini_thesis}.  
\\

\noindent
\paragraph{1. Data quality.} We perform five null tests to probe the systematic errors associated with data quality. For all we divide the data into two parts and compare the results from the two halves as described in the text.
\begin{description}
  \item[(a) Detector White Noise.]{We split data based on the white noise level in the demodulated timestreams. The split compares data above and below median white noise of the $40\,{\rm aW} \sqrt{\rm s}$.}
 \item[(b) Glitch count.]{
 We divide the data set in half based on the median value (10) of two-sample glitches in timestream for each CES and TES.}
 \item[(c) Knee frequency.] We divide the timestreams into two sets based on the knee frequency of the sum of the power spectra of demodulated 
 $Q$ and $U$ timestreams.  The median value used as the threshold is $f_{\rm knee} = 2\,{\rm mHz}$.  This assesses the performance of the HWP demodulation and the detector $1/f$ noise.
  \item[(d) Scan Synchronous Signal.] We split the data based on the amplitude of the SSS in each timestream. This data split checks the effect of scan synchronous signal (SSS) in the demodulated $Q$ and $U$ timestreams arising from structure solely dependent on azimuth, e.g., far sidelobe pick-up of ground emission. 
  \item[(e) Statistical Stationarity.]{ 
We split the data based on a statistic that assess the stationarity of the noise, a basic assumption in the analysis. The statistic is defined as fractional variation of the white noise amplitude across each CES.  The threshold for the data split is set to the median value of 3\%. 
     }
\end{description}

\noindent
\paragraph{2. Instrument performance.}
We perform six null tests to probe the systematics caused by possible malfunctioning of instrument.
\begin{description}
  \item[(a) Batch A/B detectors.]{For this test we divide the data obtained from Batch A and B detectors to capture systematics related to detector fabrication (see \S\ref{sec:instrument}).}
  \item[(b) Central/peripheral.]{We divide the data obtained from detectors that are located in the central $\theta < 8.5^{\circ}$ and 
  peripheral $\theta \ge 8.5^{\circ}$ regions of the focal plane. This test captures the systematics due to detector pointing, beam uncertainties, and performance based on their location in the focal plane. }
  \item[(c) Detector angle.]{We divide the data obtained from detectors with positive and negative polarization angles.  This test probes the systematics arising from polarization angle calibration of the detectors as well as possible systematic differences in the paired detectors sensing orthogonal polarizations. }
  \item[(d) Focal plane left/right.]{ This is similar to central/peripheral null test and probes the systematics due to difference in calibration and performance of detectors located to the left and right halves of the focal plane.}
  \item[(e) HWP performance.]{ We divide the data based on small variations in the HWP rotation. We define a figure of merit (fom) as the ratio of power in the side bands around the HWP $1f$ peak to the power in the HWP $1f$ peak. A sharper peak (low fom) represents more stable HWP rotation. We divide the data into two based on the median fom of $0.00135$. This check tests for possible sensitivity the the HWP rotation.}
  \item[(f) Readout feedback polarity.]{We divide the data from detectors with positive and negative polarity of the SQUID feedback lines.  The polarity is set for each pod.  Thus, by subtracting the maps from each feedback-line polarity, this null test probes for electrical readout noise common across the feedback loop of the 24 pods in the array.   It also somewhat randomly divides the focal plane pod-by-pod into two groups, possibly proving optical or other systematics.}
\end{description}

\noindent
\paragraph{3. Observing conditions.} We check the presence of systematics due to observing conditions by the following null tests.
\begin{description}
  \item[(a) Ambient temperature.]{ We divide the data based on the ambient temperature measurement.
              }
  \item[(b) Azimuth east/west scans.]{ We divide the data by the central azimuth. As mentioned in \S \ref{sec_observations}, ABS scanned Field A at two different azimuth centers: $125^{\circ}$ when the field is rising and $235^{\circ}$ when the field is setting. In this null test, we test the systematic that could arise due to differences in the two azimuth locations such as ground emission and pointing.}
  \item[(c)  Humidity.]{ We divide the data based on the median humidity of 19\% obtained from the APEX weather monitor.}
  \item[(d) PWV.]{ We use the median PWV value of 0.78 mm measured by the APEX radiometer to divide the dataset.}
  \item[(e) Wind speed.]{ Here we use the median wind speed of 5 km/hr to split the data into two based on APEX weather monitor data. }
\end{description}

\paragraph{4. Temporal variations and positions of the Sun and Moon.} Here we check the systematics due to temporal variations of instrument calibration and performance, as well as possible systematics due to bright sources: the Sun and Moon.
\begin{description}
\item[(a) Chronological.]{ In this test we make maps of data from the first and second halves of the full observation period before and after August 23, 2013. This test identifies systematics arising due to seasonal variations in instrument performance, calibration, and systematics such as detector responsivities and degradation of the HWP anti-reflection coating.}
\item[(b) Position of Sun and Moon.]{ We perform four null tests using the position of Sun and Moon. We make the data splits based on the elevation of the source, above and below horizon, and the angular distance between the scan center and the Sun (Moon) split at $120^\circ$ ($90^\circ$). These tests were chosen to effect of far-side lobe contamination and possible diurnal variations for the Sun related null tests.}
\end{description}

\bibliography{abs_result_v2}

\end{document}